\documentclass[aps,prb,twocolumn,superscriptaddress,nofootinbib]{revtex4-2}
\usepackage[english]{babel}
\usepackage{bm}
\usepackage{amsmath}
\usepackage{amssymb}
\usepackage{textcomp}
\usepackage{dutchcal}
\usepackage[dvips]{graphicx}
\usepackage[colorlinks=true, allcolors=cyan]{hyperref}
\usepackage{diagbox}
\usepackage{multirow}
\usepackage{threeparttable}
\usepackage{xfrac}
\usepackage{xcolor}

\begin{document}
\title{Resonant absorption and linear photovoltaic effect in ferroelectric moir\'e heterostructures}
\author{V.~V.~Enaldiev}
\email{vova.enaldiev@gmail.com}
\affiliation{Moscow Center for Advanced Studies, Kulakova str. 20, Moscow 123592, Russia} 
\affiliation{Kotelnikov Institute of Radio-engineering and Electronics of the RAS, Mokhovaya 11-7, Moscow 125009, Russia}
\affiliation{HSE University, Moscow, 101000 Russia}
\author{Z.~Z.~Alisultanov}
\email{zaur0102@gmail.com}
\affiliation{Moscow Center for Advanced Studies, Kulakova str. 20, Moscow 123592, Russia} 
\affiliation{Institute of Physics of DFRS of the RAS, Makhachkala, 367015, Russia}

\begin{abstract}
Twisted bilayers, featuring interfacial ferroelectricity in the form of array of polar domains, combined with incommensurate two-dimensional layers  in a single  van der Waals heterostructures allows for generation of purely electrostatic moir\'e superlattice potentials in the latter. We study electronic and optoelectronic properties of such heterostructures composed of graphene stacked with the twisted ferroelectric bilayers and show that doping of graphene substantially affects mini-band structures because of screening of free carriers. We demonstrate that formation of van Hove singularities in density of states modifies linear and second-order responses of the structures leading to resonant absorption and linear photovoltaic effect, respectively. The latter is generated solely by a shift photocurrent, arising only with account of virtual optical transitions, whereas an injection photocurrent is forbidden by symmetry.
\end{abstract}

\maketitle

Van der Waals (vdW) heterostructures containing moir\'e superlattices at twisted interfaces are at the forefront of solid-state physics and material science, motivated by discoveries of superconductivity in twisted WSe$_2$ \cite{Xia2024,Guo2025} and graphene \cite{cao2018unconventional,lu2019superconductors,Yankowitz2019} bilayers, strongly correlated  \cite{Wang2020,Li2021a} and other exotic \cite{bandurin2022interlayer} phases in bi- , and tri-layer \cite{Xu2021,chen2021} structures, as well as interfacial ferroelectricity in twisted hexagonal  borone nitride \cite{woods2021,stern2021,yasuda2021,Pan2025} (hBN) and transition metal dichalcogenides \cite{Weston2022,Wang2022} (TMD) bilayers. 

The interfacial ferroelectricity is a novel type of ferroelectric order specific to layered vdW materials where electric dipole moments are bounded to stacking arrangements of the layers \cite{ViznerStern2024}. In the twisted hBN and TMD heterostructures it emerges as an array of triangular domains characterized by the out-of-plane electric polarization, which scales linearly with the number of twisted interfaces \cite{Deb2022,VanWinkle2024,Ferreira2022}  and can be switched via domain wall sliding \cite{Enaldiev2022,Molino2023,Liang2023,Ko2023,Bian2024}.

Integrating an interfacial ferroelectric with a functional layer in a single vdW heterostructure enables the creation of a purely electrostatic moir\'e superlattice potential in the functional layer \cite{Magorrian2021,Zhao2021} This potential has been recently studied experimentally through electron  \cite{Ding2024,Wang2025} and exciton \cite{Kim2023} transport. A distinctive feature of such an electrostatic moir\'e potential, compared to those resulting from interlayer hybridisation in commensurate twistronic structures, is its tunability via free-carrier and proximate electrostatic gate screening as well as out-of-plane electric fields that reshape the underlying ferroelectric domains \cite{Weston2022,Enaldiev2022,Molino2023,Ko2023}. 

In the Letter, we demonstrate that an electrostatic moir\'e superlattice potential, arising from polar domains, generates mini-bands for massless Dirac fermions (mDF) in graphene with dispersion controllable by variation of carrier concentration at the fixed moir\'e superlattice period.  In addition, mini-band formation leads to a substantial enhancement of graphene's optical absorption up to $\approx10\%$ at resonant frequencies that correspond to allowed optical transitions between states at van Hove singularities (vHS). Furthermore, we show that the mini-band structure, combined with lack of inversion symmetry in the heterostructure produces linear photovoltaic effect \cite{Kraut1979,Baltz1981,morimoto2016topological,Ma2023} that is purely driven by a shift photocurrent; an injection current is absent by symmetry. Both the linear and second-order optical responses are tunable via twist angle, electron doping and an out-of-plane electric field applied across the twisted interface. 

Using periodicity of the moir\'e superlattice in the twisted bilayer we express the electrostatic moir\'e potential in graphene from polar domains as a Fourier series:
\begin{multline}\label{Eq:V_zeroD}
    V(\bm{r})=V^{(s)}_0+\\
    \sum_{\substack{l=1,2,3,\dots\\j=1,2,3}}V^{(s)}_l\cos\left(\bm{g}^{(l)}_{j}\cdot\bm{r}\right)+V^{(a)}_l\sin\left(\bm{g}^{(l)}_{j}\cdot\bm{r}\right).
\end{multline}
Here $V^{(s)}_l$ and $V^{(a)}_l$ are real Fourier amplitudes of the potential that depend on the carrier concentration in graphene, dielectric environment, and gate proximity (see section S1 in Supplementary Materials (SM)), vectors $g^{(l)}_{1,2,3}$ form the $l$-th triad of moir\'e superlattice reciprocal vectors related by $120^{\circ}$-rotation. Note, that placing the electrostatic gate near the twisted interface -- but on the side opposite to graphene layer -- doubles the electrostatic moir\'e superlattice potential in graphene compared to the case of remote gate  \cite{Magorrian2021}. For an ideal moir\'e superlattice with $120^{\circ}$-rotational symmetry, the zeroth and even-order Fourier amplitudes arise only in presence of a finite out-of-plane electric field, which creates an area imbalance between domains of opposite polarisation \cite{Weston2022,Enaldiev2022,Ko2023}.

\begin{figure*}
	\includegraphics[width=2.0\columnwidth]{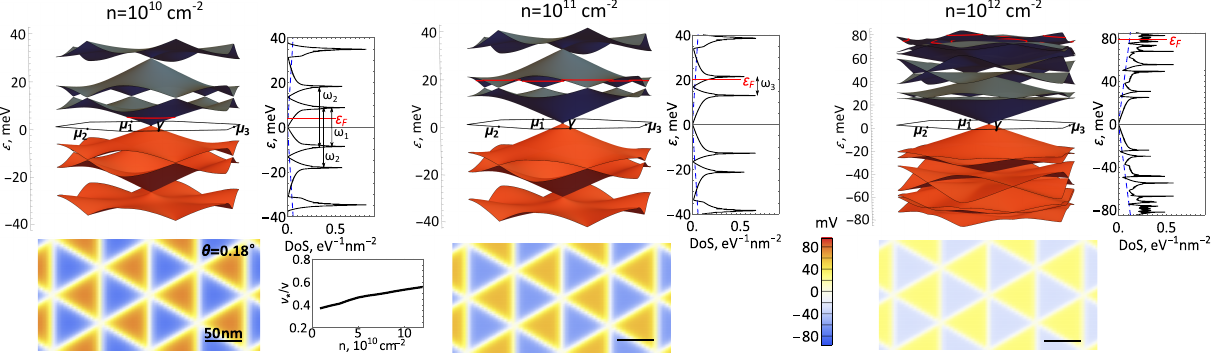}
	\caption{\label{fig:minibands} Low-energy mini-bands and corresponding densities of states for graphene/twisted hBN bilayer heterostructure characterized by twist angle $\theta=0.18^{\circ}$ for three values of electron doping $n=10^{10}$, $10^{11}$ and $10^{12}$ cm$^{-2}$ (red lines shows corresponding Fermi-energies $\varepsilon_F$). Blue dashed line shows density of states in an isolated graphene; arrows indicate resonant frequencies in absorption. Bottom insets show distribution of electrostatic moir\'e potential in graphene created by polar domains in twisted hBN bilayer and renormalization of primary mDF group velocity with electron concentration at $\theta=0.18^{\circ}$ . }
\end{figure*}

We begin with calculation of a mini-band structure for free carriers in graphene with potential \eqref{Eq:V_zeroD} which reduces to solution of the following equation  \cite{CastroNeto2009,fn1}:
\begin{align}
    \hat{H}\psi&=\varepsilon\psi, \label{Eq:Weyl}\\    
    \hat{H}&=v\bm{\sigma}\cdot\hat{\bm{p}}+V(\bm{r}), \label{Eq:Ham}
\end{align}
where $\psi=(\psi_1,\psi_2)^{\rm T}$ and $v\approx 10^6$\,m/s are a two-component envelope wave function and the speed of Dirac fermions in graphene, respectively, $\bm{\sigma}=( \sigma_x,\sigma_y)$ is 2D vector of the Pauli matrices acting in sublattice space, $\hat{\bm{p}}=-i\hbar(\partial/\partial x,\partial/\partial y)$ is the quasi-momentum operator. Eq. \eqref{Eq:Weyl} can be diagonalized in the basis of Bloch functions $\psi_{n,\bm{k}}=e^{i\bm{q}\bm{r}}u_{n,\bm{k}}(\bm{r})$ corresponding to eigen values $\varepsilon_n(\bm{k})$, where $u_{n,\bm{k}}(\bm{r})$ is a moir\'e periodic Bloch amplitude characterized by the miniband index, $n=1,2,3,\dots$ for electrons  ($n=-1,-2,-3,\dots$ for holes), and a quasi-momentum, $\bm{k}$, defined in moir\'e Brillouin zone (mBZ).

Since an account of finite out-of-plane electric field does not change results obtained for zero electric field, in the main text  we focus on the latter case. The corresponding results for a finite field are gathered in sections S4 and S5 of SM. In the absence of the out-of-plane electric field the moir\'e  potential \eqref{Eq:V_zeroD} possesses odd parity, $V(-\bm{r})=-V(\bm{r})$, illustrated on the bottom insets in Fig. \ref{fig:minibands} ($V^{(s)}_{l}\equiv0$, $l=0,1,2,\dots$  ). The operator $\hat{I}$, transforming $\bm{r}\to-\bm{r}$, then, corresponds to the chiral symmetry of Hamiltonian \eqref{Eq:Ham}, satisfying $\hat{H}\hat{I}=-\hat{I}\hat{H}$. Combined with an intravalley hidden symmetry $\hat{T}=i\sigma_y\hat{K}$ (where $\hat{K}$ is the complex conjugation operator and $\hat{H}\hat{T}=\hat{T}\hat{H}$), this leads to electron-hole symmetry of states with the same quasi-momentum: $\varepsilon_{n}(\bm{k})=-\varepsilon_{-n}(\bm{k})$. We also note that the Hamiltonian  \eqref{Eq:Ham} is invariant under non-centrosymmetric $C_{3v}$ point group, which includes, in particular, mirror reflection with respect to the $y$-axis, represented by an operator, $\hat{M}_y=\sigma_x\{y\to-y\}$ used below.   

In Fig. \ref{fig:minibands} we demonstrate the dispersions obtained for several values of graphene doping, calculated with the help of expansion of $u_{n,\bm{k}}(\bm{r})$ in Fourier series and diagonalization of the resulting matrix for all quasimomenta $\bm{k}$ in mBZ, with the number of harmonics large enough to satisfy convergence in the low-energy spectral interval.  The lowest mini-bands ($n=\pm1$) are characterised by gapless spectrum at the neutrality point ($\varepsilon=0$) as the electrostatic moir\'e potential does not break the sublattice symmetry required to produce a gap for mDF. However, the group velocity of $n=\pm1$ minibands  can be substantially suppressed by the moir\'e potential. To demonstrate this we project out Eq. \eqref{Eq:V_zeroD} to a subspace of the two degenerate states  $\psi_{\pm1,\bm{k}=0}$ and perturbatively account for $v\,\bm{\sigma}\cdot\bm{k}$. Then, leaving in Fourier series for $u_{\pm1,\bm{k}=0}$ only the zeroth and $\pm\bm{g}^{(1)}_{1,2,3}$ harmonics, we obtain the following effective Hamiltonian (see section S2 in SM):
\begin{equation}
    H_{\bm{\gamma}}=\hbar v_*\bm{\sigma}\cdot\bm{k},
\end{equation}
characterised by the renormalized group velocity
\begin{equation}\label{Eq:v*}
    v_*=\frac{v}{1+\left(\frac{3V_1^{(a)} a}{4\sqrt{2}\pi\hbar v \theta }\right)^2}.
\end{equation} 
From Eq. \eqref{Eq:v*} it follows that $v_*$ is a funciton of both moir\'e superlattice period ($\approx a/\theta$) and, implicitly, free carrier density via  $V_1^{(a)}$. Unlike graphene/hBN moir\'e superlattices \cite{Ortix2012,Wallbank2013} or periodically patterned graphene \cite{Park2008a,Park2008} where renormalization of the primary mDF velocity is small, here, decrease of $v_*$ can be as large as $\approx 3$ times for realistic parameters (see bottom inset in Fig. \ref{fig:minibands}). Opportunity to tune the mDF velocity can have practical significance for adjusting transmission probability of p-n junctions \cite{Cheianov2006} in the heterostructures, whereas its reduction  may help to observe Landau level breakdown even at moderate homogeneous electric fields \cite{Lukose2007,alisultanov2014landau}. 

At  $\bm{k}=\bm{g}^{(1)}_{1,2,3}/2\equiv\bm{\mu}_{1,2,3}$ located at the mBZ edges, pairs of electron ($n=1$ and $n=2$) and hole ($n=-1$ and $n=-2$) mini-bands become degenerate, forming secondary Dirac points symmetrically with respect to $\varepsilon=0$. Around the degeneracy momenta, $\bm{q}=\bm{k}-\bm{\mu}_{i}$, an effective model can be constructed accounting for coupling of the two harmonics $|\bm{\mu}_{i}+\bm{q}\rangle$ and $|\bm{\mu}_i+\bm{q}-\bm{g}_i\rangle$ in the Fourier expansion of $u_{\bm{\mu}_i+\bm{q}}$, which leads to \cite{fn2} (see section S3 in SM): 
\begin{eqnarray}\label{Eq:Ham_mu}
    H^{(\pm)}_{\bm{\mu}_3} = \qquad\qquad\qquad\qquad\qquad\qquad\qquad\qquad\qquad\\ 
    \pm\left(\varepsilon_{\bm{\mu}}+\frac{(\hbar v_*)^4g^2}{8\varepsilon^3_{\bm{\mu}}}q_y^2\right)\sigma_0\pm\frac{V_1^{(a)}\hbar v_*}{2\varepsilon_{\bm{\mu}}}q_y\sigma_z-\hbar v_*q_x\sigma_x, \nonumber
\end{eqnarray}
where $+/-$ sign is for  electrons/holes, $\pm\varepsilon_{\bm{\mu}}=\pm\sqrt{(\hbar v_*g)^2+\left(V_1^{(a)}\right)^2}/2$ are energies of the secondary Dirac points ($g=|\bm{g}_1^{1}|$), and $\sigma_0$ is an identity matrix. Hamiltonian \eqref{Eq:Ham_mu} gives rise to strongly anisotropic mDF (see Fig. \ref{fig:minibands}), where in contrast to graphene/hBN \cite{Ortix2012,Wallbank2013,Yankowitz2012} or patterned graphene \cite{Park2008a}  velocities in both principal axes can be controlled by carrier concentration and twist angle. 

Increasing the doping level, the Fermi contour undergo Lifshitz transition shifting from a central $\gamma$ pocket to those in $\mu_{1,2,3}$ followed by the formation of a pair of vHS in density of states (DOS). We reveal below that the formation of these electron-hole symmetric vHS significantly modifies both the linear and second-order responses of the system in infrared energy range. 

We consider normal incidence of a plane monochromatic electromagnetic wave, described by vector-potential $\bm{A}(\bm{r},t)=A_0\bm{\eta}\cos(kz-\omega t)$, where $A_0$ and $\bm{\eta}=(\cos\chi,\sin\chi,0)$ characterize the amplitude and polarisation of the wave, respectively. Using first-order perturbation theory, the interband absorption coefficient is expressed as follows \cite{Callaway1964} (in SI units): 
\begin{align}\label{Eq:absorption}
\alpha(\omega)&= \frac{g_s g_v\pi e^2}{c\omega\epsilon_0 S}\times\\ 
      \sum_{n,n',\bm{k}}&\left|(\hat{\bm{v}}_{nn'}(\bm{k})\cdot\bm{\eta})\right|^2\left[f_{n',\bm{k}}-f_{n,\bm{k}}\right]\delta\left(\varepsilon_{n,\bm{k}}-\varepsilon_{n',\bm{k}}-\hbar\omega\right),     \nonumber
\end{align}
where $g_s=2$ and $g_v=2$ are the spin and valley degeneracy factors in graphene, $c$ is speed of light, $\epsilon_0$ is the dielectric permittivity of vacuum, $S$ is the area of the structure, $n,n'$ are the mini-band indexes, $\hat{\bm{v}}_{nn'}(\bm{k})$ is interband matrix element of velocity operator $\hat{\bm{v}}=\partial H/\partial\hat{p} =v\bm{\sigma}$ on the Bloch states characterised by quasi-momentum $\bm{k}$,  and $f_{n/n',\bm{k}}\equiv f(\varepsilon_{n/n',\bm{k}})$ are the Fermi-Dirac distribution functions characterised by  equilibrium for given electron density. 

\begin{figure}[!t]
	
	\includegraphics[width=1.0\columnwidth]{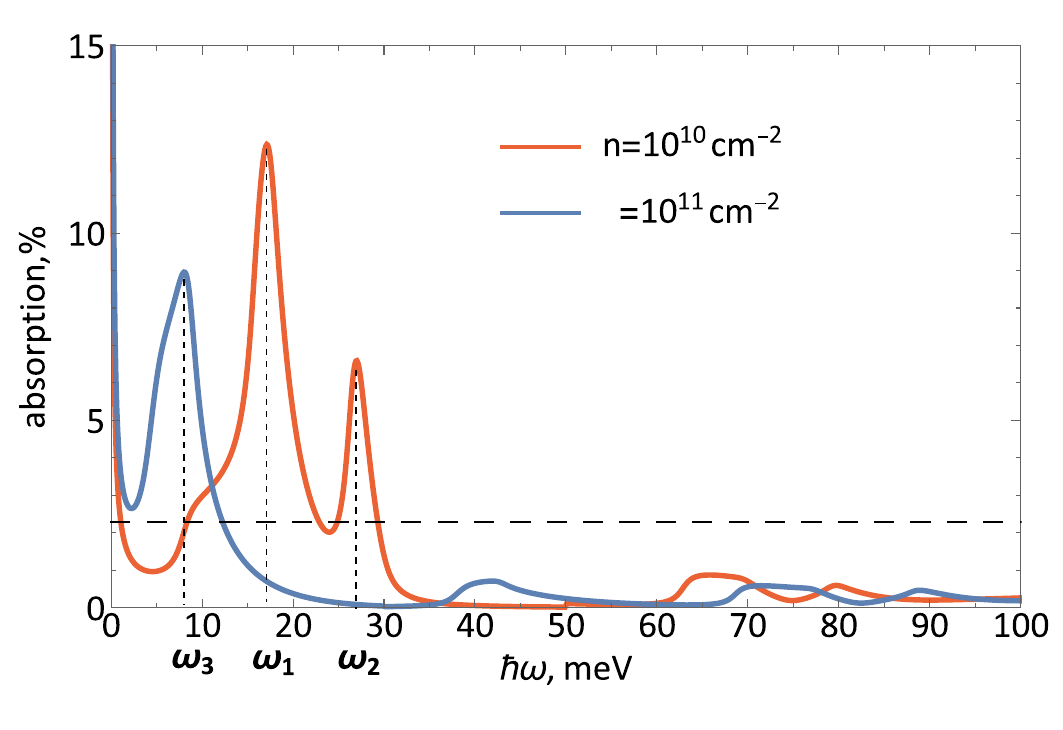}
	\caption{\label{fig:absorption} Frequency dependences of the absorption coefficient \eqref{Eq:absorption} for doping levels corresponding to Fig. \ref{fig:minibands} at $T\ll\hbar\omega_3$. The resonances, $\omega_{1,2,3}$, corresponds to energy differences between states at peaks of DOS, shown in Fig. \ref{fig:minibands}, where the interband velocity matrix elements are non-zero. Dashed line indicates absorption ($2.3\%$) of an isolated graphene layer \cite{Nair2008}.}
\end{figure}

 The $\alpha(\omega)$ dependences, displayed in Fig. \ref{fig:absorption} for $\theta=0.18^{\circ}$, exhibit resonances whose frequencies depend on the filling of mini-band structure. At an electron concentration of $n=10^{10}$\,cm$^{-2}$ (corresponding to filling of $n=1$ mini-band before the Lifshitz transition, left panel in Fig. \ref{fig:minibands}), the absorption coefficient is dominated by three terms in Eq. \eqref{Eq:absorption}: $(n',n)=(-1,1)$, $(-1,2)$ and $(-2,1)$. The first term accounts for the peak at $\omega_1$, while the latter two contribute equally to the peak at  $\omega_2$. 
 
 At a higher concentration, $n=10^{11}$\,cm$^{-2}$ (middle panel in Fig. \ref{fig:minibands}), transitions between electron and hole mini-bands are suppressed by Pauli blocking, whereas transition between electron  $n'=1$ and $n=2$ mini-bands become activated in Eq. \eqref{Eq:absorption}, resulting in a peak at frequency $\omega_3$. In both cases, the resonances arise from allowed optical transitions from filled states at one vHS to empty states of another vHS forming close to the boundary of mid- and far-infrared range. 
 
 We note that the magnitude of absorption at these resonances can significantly exceed that of an individual graphene layer \cite{Nair2008} (dashed line in Fig. \ref{fig:absorption}), making it competitive with plasmon absorption of graphene nanostructures \cite{Ju2011}.  
 
To observe the resonant absorption, the temperature should satisfy  $T\ll\hbar\omega_{1,2,3}$ to ensure complete population and depletion of the corresponding vHS responsible for each resonance. For the case of $\theta=0.18^{\circ}$, shown in Fig. \ref{fig:absorption}, this requires cryogenic temperatures for every resonance. However, since quasi-momentum states of $n=\pm1,\pm2$ minibands, that form vHS, lie near the mBZ edge, the resonance frequencies -- particularly $\omega_{1,2}$ -- can be readily tuned by adjusting $\theta$. This allows the condition to be met even at room temperatures for larger twist angles, with the shift of resonance frequency deeper into the mid-infrared range. 

\begin{figure}
	\includegraphics[width=1.0\columnwidth]{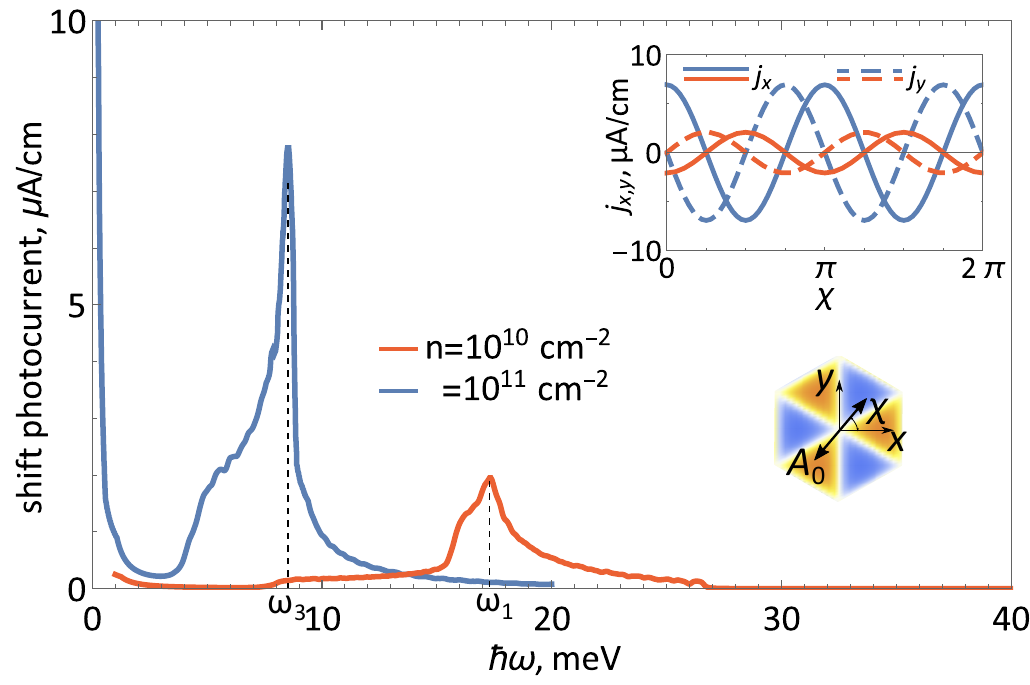}
	\caption{\label{fig:photocurrent}Frequency dependences of the shift photocurrent \eqref{Eq:jshift} for graphene--twisted hBN bilayer with $\theta=0.18^{\circ}$ at the intensity of the incident radiation $I=0.1 {\rm W/cm^2}$ and indicated electron dopings. Insets: (Top)  Dependences of  $j^{x,y}_{\rm shift}(\omega_{1},\chi)$ (red) and $j^{x,y}_{\rm shift}(\omega_{3},\chi)$ (blue) on the polarisation of the incident electromagnetic wave, characterised by angle $\chi$. (Bottom) Orientation of the polarisation with respect to moir\'e superlattice in the twisted bilayer.} 
\end{figure}

Now, we consider second order response and show the emergence of the linear photovoltaic effect \cite{Kraut1979,Baltz1981} associated with resonant shift photocurrent. To this end, solving  the quantum kinetic equation (see details in SM), we find second order amendment to the single-particle density matrix on the incident wave amplitude, $\rho^{(2)}\propto A_0^2$, and then, calculate photocurrent as follows: 
\begin{eqnarray}\label{Eq:photocurrent}
    \bm{j}=\frac{1}{S}{\rm Tr} (e\hat{\bm{v}}\rho^{(2)}) = \bm{j}_{\rm inj}+\bm{j}_{\rm shift}.
\end{eqnarray}
Here ${\rm Tr}$ denotes operator trace, the first term on the right hand side describes an injection photocurrent
\begin{multline}\label{Eq:jinj}
     \bm{j}_{\rm inj} =\frac{g_sg_v\pi e^3I\tau}{\hbar c\epsilon_0\omega^2S}\sum_{\substack{n,n'\\\bm{k},s=\pm}}\left[f_{n'}-f_{n}\right]\delta\left(\varepsilon_{n}-\varepsilon_{n'}-s\hbar\omega\right)\times\\
     \times\hat{\bm{v}}_{nn} (\hat{\bm{v}}_{nn'}\cdot\bm{\eta})(\hat{\bm{v}}_{n'n}\cdot\bm{\eta}),
\end{multline}
whereas the second is a shift photocurrent
\begin{multline}\label{Eq:jshift}
     \bm{j}_{\rm shift} =\frac{g_sg_v\pi e^3I}{c\epsilon_0\omega^2 S}\sum_{\substack{n,n',n''\neq n'\\\bm{k},s=\pm}}\left[f_{n'}-f_{n}\right]\delta\left(\varepsilon_{n}-\varepsilon_{n'}-s\hbar\omega\right)\times\\
     \times{\rm Im}\left[\frac{\hat{\bm{v}}_{n'n''} (\hat{\bm{v}}_{n''n}\cdot\bm{\eta})(\hat{\bm{v}}_{nn'}\cdot\bm{\eta})}{\varepsilon_{n''}-\varepsilon_{n'}}\right].
\end{multline}
In Eqs. \eqref{Eq:jinj} and \eqref{Eq:jshift} $I=c\epsilon_0\omega^2A_0^2/2$ is the incident wave intensity, $\tau$ is a phenomenological momentum relaxation time and  $\bm{k}$ is suppressed in arguments of matrix elements and energy values.  The hidden intravalley symmetry,  $\hat{T}$  of the Hamiltonian \eqref{Eq:Weyl} leads to the following relation \\\cite{fn3}: 
\begin{equation}\label{Eq:Tsym}
    \bm{v}_{nn'}(\bm{k}) = -e^{i\left[\alpha_{n'}(\bm{k})-\alpha_{n}(\bm{k})\right]} \bm{v}_{nn'}^*(-\bm{k}), 
\end{equation}
which has already been used in derivation of Eq. \eqref{Eq:jshift} ($\alpha_{n'}(\bm{k})$ and $\alpha_{n}(\bm{k})$ are arbitrary gauge phases). Self-conjugation of $\hat{\bm{v}}$ supplemented with Eq. \eqref{Eq:Tsym} and $\varepsilon_{n,\bm{k}}=\varepsilon_{n,-\bm{k}}$  leads to vanishing of the injection current \eqref{Eq:jinj} independently on the polarisation of the wave. We note that account of small trigonal warping terms in Hamiltonian \eqref{Eq:Weyl} would only lead to a valley-polarised injection photocurrent as in isolated graphene layer \cite{Golub2011}, whereas a net contribution of $+{\rm K}$ and $-{\rm K}$ valleys would compensate each other \cite{fn4}.
 
Nevertheless, the shift photocurrent  \eqref{Eq:jshift} is non-zero and exhibits resonances at $\omega_1$ or $\omega_3$ depending on the carrier density, as shown in Fig. \ref{fig:photocurrent}. However, the shape of the resonances as a function of frequency differs from that of the absorption coefficient. The latter arises from real interband transitions between states that satisfy energy conservation, $\varepsilon_{n,\bm{k}}-\varepsilon_{n',\bm{k}}-\hbar\omega$. In contrast, the shift photocurrent relies crucially on virtual interband transitions, which do not require energy conversation for the intermediate states. 

Indeed, by restricting the summation in Eq. \eqref{Eq:jshift} to $n$ and $n'$ bands satisfying energy conservation in the delta-function (i.e. retaining only terms with $n''=n$ for arbitrary $n$, $n'$) and, for definiteness, specifying polarisation  $\bm{\eta}\,||\,Ox$ we obtain
\begin{align}
    j^x_{\rm shift}=& 
    \sum_{\bm{k}}F(\varepsilon_n,\varepsilon_n'){\rm Im}\left[\left|\hat{v}^x_{n'n}(\bm{k})\right|^2\hat{v}_{nn}(\bm{k})\right]=0, \label{Eq:jx}\\
    j^y_{\rm shift}=&\sum_{\bm{k}}F(\varepsilon_n,\varepsilon_n'){\rm Im}\left[\hat{v}^y_{n'n}(\bm{k})\hat{v}^x_{nn'}(\bm{k})\hat{v}^x_{nn}(\bm{k})\right]=0,    \label{Eq:jy}
\end{align}
where $F(\varepsilon_n,\varepsilon_n')$ contains only energy-dependent factors from Eq. \eqref{Eq:jshift} invariant under $C_{3v}$ point group symmetry. Therefore, the vanishing of $j^x_{\rm shift}$ \eqref{Eq:jx}  occurs because group velocity $\hat{v}_{nn}(\bm{k})=\partial \varepsilon_{n}(\bm{k})/\partial \bm{k}$ is real. On the other hand, mirror symmetry, $\hat{M}_y$, of the Hamiltonian \eqref{Eq:Weyl} leads to $\hat{v}^y_{n'n}(k_x,-k_y)=-\hat{v}^y_{n'n}(k_x,k_y)$, $\hat{v}^x_{n'n}(k_x,-k_y)=\hat{v}^x_{n'n}(k_x,k_y)$, equalizing the sum over $\bm{k}$  in Eq. \eqref{Eq:jy} to zero.  

These arguments demonstrate that the existence of more than two bands coupled via velocity matrix elements is a necessary condition for emergence of the shift photocurrent in the heterostructure, which is a distinctive feature of the photovoltaic effect in pure crystals \cite{Kraut1979,Baltz1981}.

On the inset in Fig. \ref{fig:photocurrent} we show polarisation dependence of the photocurrent at resonant frequencies, $\bm{j}_{\rm shift}(\chi)\propto(\cos 2\chi,-\sin2\chi)$ specific for $C_{3v}$ point group \cite{fn5}. 

Finally, we discuss the effect of an out-of-plane electric field across the twisted bilayer.  While such a field breaks the chiral symmetry of $\hat{H}$ by modifying the domain structure \cite{Weston2022,Enaldiev2022}, it does not eliminate van Hove singularities in density of states. Therefore, the resonant responses persists at finite out-of-plane electric field that provides an additional external tool to control the resonant conditions. Moreover, since the electric field preserves $\hat{T}$ and the spatial symmetries, the photovoltaic effect remains governed by the shift photocurrent, while the injection current also vanishes.  

We believe that the resonant linear photovoltaic effect should not be exclusive to moir\'e potential of polar domains in graphene, but is also likely present in bilayer graphene and within the intrinsic electron/hole mini-bands of twisted TMD bilayers characterized by broken inversion symmetry and vHS \cite{Magorrian2021}. 
 
{\it Acknowledgements.} We thank Denis Bandurin for fruitful discussions. The work was supported by the Russian Science Foundation (main text) (project No. 24-72-10015) and Ministry of Science and Higher Education
of the Russian Federation (results for finite out-of-plane electric field in Supplementary Materials) (No. FSMG-2026-0012).


\bibliography{refer}

\begin{thebibliography}{56}%
\makeatletter
\providecommand \@ifxundefined [1]{%
 \@ifx{#1\undefined}
}%
\providecommand \@ifnum [1]{%
 \ifnum #1\expandafter \@firstoftwo
 \else \expandafter \@secondoftwo
 \fi
}%
\providecommand \@ifx [1]{%
 \ifx #1\expandafter \@firstoftwo
 \else \expandafter \@secondoftwo
 \fi
}%
\providecommand \natexlab [1]{#1}%
\providecommand \enquote  [1]{``#1''}%
\providecommand \bibnamefont  [1]{#1}%
\providecommand \bibfnamefont [1]{#1}%
\providecommand \citenamefont [1]{#1}%
\providecommand \href@noop [0]{\@secondoftwo}%
\providecommand \href [0]{\begingroup \@sanitize@url \@href}%
\providecommand \@href[1]{\@@startlink{#1}\@@href}%
\providecommand \@@href[1]{\endgroup#1\@@endlink}%
\providecommand \@sanitize@url [0]{\catcode `\\12\catcode `\$12\catcode
  `\&12\catcode `\#12\catcode `\^12\catcode `\_12\catcode `\%12\relax}%
\providecommand \@@startlink[1]{}%
\providecommand \@@endlink[0]{}%
\providecommand \url  [0]{\begingroup\@sanitize@url \@url }%
\providecommand \@url [1]{\endgroup\@href {#1}{\urlprefix }}%
\providecommand \urlprefix  [0]{URL }%
\providecommand \Eprint [0]{\href }%
\providecommand \doibase [0]{https://doi.org/}%
\providecommand \selectlanguage [0]{\@gobble}%
\providecommand \bibinfo  [0]{\@secondoftwo}%
\providecommand \bibfield  [0]{\@secondoftwo}%
\providecommand \translation [1]{[#1]}%
\providecommand \BibitemOpen [0]{}%
\providecommand \bibitemStop [0]{}%
\providecommand \bibitemNoStop [0]{.\EOS\space}%
\providecommand \EOS [0]{\spacefactor3000\relax}%
\providecommand \BibitemShut  [1]{\csname bibitem#1\endcsname}%
\let\auto@bib@innerbib\@empty
\bibitem [{\citenamefont {Xia}\ \emph {et~al.}(2024)\citenamefont {Xia},
  \citenamefont {Han}, \citenamefont {Watanabe}, \citenamefont {Taniguchi},
  \citenamefont {Shan},\ and\ \citenamefont {Mak}}]{Xia2024}%
  \BibitemOpen
  \bibfield  {author} {\bibinfo {author} {\bibfnamefont {Y.}~\bibnamefont
  {Xia}}, \bibinfo {author} {\bibfnamefont {Z.}~\bibnamefont {Han}}, \bibinfo
  {author} {\bibfnamefont {K.}~\bibnamefont {Watanabe}}, \bibinfo {author}
  {\bibfnamefont {T.}~\bibnamefont {Taniguchi}}, \bibinfo {author}
  {\bibfnamefont {J.}~\bibnamefont {Shan}},\ and\ \bibinfo {author}
  {\bibfnamefont {K.~F.}\ \bibnamefont {Mak}},\ }\bibfield  {title} {\bibinfo
  {title} {Superconductivity in twisted bilayer wse2},\ }\href
  {https://doi.org/10.1038/s41586-024-08116-2} {\bibfield  {journal} {\bibinfo
  {journal} {Nature}\ }\textbf {\bibinfo {volume} {637}},\ \bibinfo {pages}
  {833} (\bibinfo {year} {2024})}\BibitemShut {NoStop}%
\bibitem [{\citenamefont {Guo}\ \emph {et~al.}(2025)\citenamefont {Guo},
  \citenamefont {Pack}, \citenamefont {Swann}, \citenamefont {Holtzman},
  \citenamefont {Cothrine}, \citenamefont {Watanabe}, \citenamefont
  {Taniguchi}, \citenamefont {Mandrus}, \citenamefont {Barmak}, \citenamefont
  {Hone}, \citenamefont {Millis}, \citenamefont {Pasupathy},\ and\
  \citenamefont {Dean}}]{Guo2025}%
  \BibitemOpen
  \bibfield  {author} {\bibinfo {author} {\bibfnamefont {Y.}~\bibnamefont
  {Guo}}, \bibinfo {author} {\bibfnamefont {J.}~\bibnamefont {Pack}}, \bibinfo
  {author} {\bibfnamefont {J.}~\bibnamefont {Swann}}, \bibinfo {author}
  {\bibfnamefont {L.}~\bibnamefont {Holtzman}}, \bibinfo {author}
  {\bibfnamefont {M.}~\bibnamefont {Cothrine}}, \bibinfo {author}
  {\bibfnamefont {K.}~\bibnamefont {Watanabe}}, \bibinfo {author}
  {\bibfnamefont {T.}~\bibnamefont {Taniguchi}}, \bibinfo {author}
  {\bibfnamefont {D.~G.}\ \bibnamefont {Mandrus}}, \bibinfo {author}
  {\bibfnamefont {K.}~\bibnamefont {Barmak}}, \bibinfo {author} {\bibfnamefont
  {J.}~\bibnamefont {Hone}}, \bibinfo {author} {\bibfnamefont {A.~J.}\
  \bibnamefont {Millis}}, \bibinfo {author} {\bibfnamefont {A.}~\bibnamefont
  {Pasupathy}},\ and\ \bibinfo {author} {\bibfnamefont {C.~R.}\ \bibnamefont
  {Dean}},\ }\bibfield  {title} {\bibinfo {title} {Superconductivity in 5.0°
  twisted bilayer wse2},\ }\href {https://doi.org/10.1038/s41586-024-08381-1}
  {\bibfield  {journal} {\bibinfo  {journal} {Nature}\ }\textbf {\bibinfo
  {volume} {637}},\ \bibinfo {pages} {839} (\bibinfo {year}
  {2025})}\BibitemShut {NoStop}%
\bibitem [{\citenamefont {Cao}\ \emph {et~al.}(2018)\citenamefont {Cao},
  \citenamefont {Fatemi}, \citenamefont {Fang}, \citenamefont {Watanabe},
  \citenamefont {Taniguchi}, \citenamefont {Kaxiras},\ and\ \citenamefont
  {Jarillo-Herrero}}]{cao2018unconventional}%
  \BibitemOpen
  \bibfield  {author} {\bibinfo {author} {\bibfnamefont {Y.}~\bibnamefont
  {Cao}}, \bibinfo {author} {\bibfnamefont {V.}~\bibnamefont {Fatemi}},
  \bibinfo {author} {\bibfnamefont {S.}~\bibnamefont {Fang}}, \bibinfo {author}
  {\bibfnamefont {K.}~\bibnamefont {Watanabe}}, \bibinfo {author}
  {\bibfnamefont {T.}~\bibnamefont {Taniguchi}}, \bibinfo {author}
  {\bibfnamefont {E.}~\bibnamefont {Kaxiras}},\ and\ \bibinfo {author}
  {\bibfnamefont {P.}~\bibnamefont {Jarillo-Herrero}},\ }\bibfield  {title}
  {\bibinfo {title} {Unconventional superconductivity in magic-angle graphene
  superlattices},\ }\href {https://doi.org/10.1038/nature26160} {\bibfield
  {journal} {\bibinfo  {journal} {Nature}\ }\textbf {\bibinfo {volume} {556}},\
  \bibinfo {pages} {43} (\bibinfo {year} {2018})}\BibitemShut {NoStop}%
\bibitem [{\citenamefont {Lu}\ \emph {et~al.}(2019)\citenamefont {Lu},
  \citenamefont {Stepanov}, \citenamefont {Yang}, \citenamefont {Xie},
  \citenamefont {Aamir}, \citenamefont {Das}, \citenamefont {Urgell},
  \citenamefont {Watanabe}, \citenamefont {Taniguchi}, \citenamefont {Zhang}
  \emph {et~al.}}]{lu2019superconductors}%
  \BibitemOpen
  \bibfield  {author} {\bibinfo {author} {\bibfnamefont {X.}~\bibnamefont
  {Lu}}, \bibinfo {author} {\bibfnamefont {P.}~\bibnamefont {Stepanov}},
  \bibinfo {author} {\bibfnamefont {W.}~\bibnamefont {Yang}}, \bibinfo {author}
  {\bibfnamefont {M.}~\bibnamefont {Xie}}, \bibinfo {author} {\bibfnamefont
  {M.~A.}\ \bibnamefont {Aamir}}, \bibinfo {author} {\bibfnamefont
  {I.}~\bibnamefont {Das}}, \bibinfo {author} {\bibfnamefont {C.}~\bibnamefont
  {Urgell}}, \bibinfo {author} {\bibfnamefont {K.}~\bibnamefont {Watanabe}},
  \bibinfo {author} {\bibfnamefont {T.}~\bibnamefont {Taniguchi}}, \bibinfo
  {author} {\bibfnamefont {G.}~\bibnamefont {Zhang}}, \emph {et~al.},\
  }\bibfield  {title} {\bibinfo {title} {Superconductors, orbital magnets and
  correlated states in magic-angle bilayer graphene},\ }\href
  {https://doi.org/10.1038/s41586-019-1695-0} {\bibfield  {journal} {\bibinfo
  {journal} {Nature}\ }\textbf {\bibinfo {volume} {574}},\ \bibinfo {pages}
  {653} (\bibinfo {year} {2019})}\BibitemShut {NoStop}%
\bibitem [{\citenamefont {{Yankowitz}}\ \emph {et~al.}(2019)\citenamefont
  {{Yankowitz}}, \citenamefont {{Chen}}, \citenamefont {{Polshyn}},
  \citenamefont {{Zhang}}, \citenamefont {{Watanabe}}, \citenamefont
  {{Taniguchi}}, \citenamefont {{Graf}}, \citenamefont {{Young}},\ and\
  \citenamefont {{Dean}}}]{Yankowitz2019}%
  \BibitemOpen
  \bibfield  {author} {\bibinfo {author} {\bibfnamefont {M.}~\bibnamefont
  {{Yankowitz}}}, \bibinfo {author} {\bibfnamefont {S.}~\bibnamefont {{Chen}}},
  \bibinfo {author} {\bibfnamefont {H.}~\bibnamefont {{Polshyn}}}, \bibinfo
  {author} {\bibfnamefont {Y.}~\bibnamefont {{Zhang}}}, \bibinfo {author}
  {\bibfnamefont {K.}~\bibnamefont {{Watanabe}}}, \bibinfo {author}
  {\bibfnamefont {T.}~\bibnamefont {{Taniguchi}}}, \bibinfo {author}
  {\bibfnamefont {D.}~\bibnamefont {{Graf}}}, \bibinfo {author} {\bibfnamefont
  {A.~F.}\ \bibnamefont {{Young}}},\ and\ \bibinfo {author} {\bibfnamefont
  {C.~R.}\ \bibnamefont {{Dean}}},\ }\bibfield  {title} {\bibinfo {title}
  {{Tuning superconductivity in twisted bilayer graphene}},\ }\href
  {https://doi.org/10.1126/science.aav1910} {\bibfield  {journal} {\bibinfo
  {journal} {Science}\ }\textbf {\bibinfo {volume} {363}},\ \bibinfo {pages}
  {1059} (\bibinfo {year} {2019})}\BibitemShut {NoStop}%
\bibitem [{\citenamefont {Wang}\ \emph {et~al.}(2020)\citenamefont {Wang},
  \citenamefont {Shih}, \citenamefont {Ghiotto}, \citenamefont {Xian},
  \citenamefont {Rhodes}, \citenamefont {Tan}, \citenamefont {Claassen},
  \citenamefont {Kennes}, \citenamefont {Bai}, \citenamefont {Kim},
  \citenamefont {Watanabe}, \citenamefont {Taniguchi}, \citenamefont {Zhu},
  \citenamefont {Hone}, \citenamefont {Rubio}, \citenamefont {Pasupathy},\ and\
  \citenamefont {Dean}}]{Wang2020}%
  \BibitemOpen
  \bibfield  {author} {\bibinfo {author} {\bibfnamefont {L.}~\bibnamefont
  {Wang}}, \bibinfo {author} {\bibfnamefont {E.-M.}\ \bibnamefont {Shih}},
  \bibinfo {author} {\bibfnamefont {A.}~\bibnamefont {Ghiotto}}, \bibinfo
  {author} {\bibfnamefont {L.}~\bibnamefont {Xian}}, \bibinfo {author}
  {\bibfnamefont {D.~A.}\ \bibnamefont {Rhodes}}, \bibinfo {author}
  {\bibfnamefont {C.}~\bibnamefont {Tan}}, \bibinfo {author} {\bibfnamefont
  {M.}~\bibnamefont {Claassen}}, \bibinfo {author} {\bibfnamefont {D.~M.}\
  \bibnamefont {Kennes}}, \bibinfo {author} {\bibfnamefont {Y.}~\bibnamefont
  {Bai}}, \bibinfo {author} {\bibfnamefont {B.}~\bibnamefont {Kim}}, \bibinfo
  {author} {\bibfnamefont {K.}~\bibnamefont {Watanabe}}, \bibinfo {author}
  {\bibfnamefont {T.}~\bibnamefont {Taniguchi}}, \bibinfo {author}
  {\bibfnamefont {X.}~\bibnamefont {Zhu}}, \bibinfo {author} {\bibfnamefont
  {J.}~\bibnamefont {Hone}}, \bibinfo {author} {\bibfnamefont {A.}~\bibnamefont
  {Rubio}}, \bibinfo {author} {\bibfnamefont {A.~N.}\ \bibnamefont
  {Pasupathy}},\ and\ \bibinfo {author} {\bibfnamefont {C.~R.}\ \bibnamefont
  {Dean}},\ }\bibfield  {title} {\bibinfo {title} {Correlated electronic phases
  in twisted bilayer transition metal dichalcogenides},\ }\href
  {https://doi.org/10.1038/s41563-020-0708-6} {\bibfield  {journal} {\bibinfo
  {journal} {Nature Materials}\ }\textbf {\bibinfo {volume} {19}},\ \bibinfo
  {pages} {861} (\bibinfo {year} {2020})}\BibitemShut {NoStop}%
\bibitem [{\citenamefont {Li}\ \emph {et~al.}(2021)\citenamefont {Li},
  \citenamefont {Hu}, \citenamefont {Feng}, \citenamefont {Zhou}, \citenamefont
  {An}, \citenamefont {Law}, \citenamefont {Wang},\ and\ \citenamefont
  {Lin}}]{Li2021a}%
  \BibitemOpen
  \bibfield  {author} {\bibinfo {author} {\bibfnamefont {E.}~\bibnamefont
  {Li}}, \bibinfo {author} {\bibfnamefont {J.-X.}\ \bibnamefont {Hu}}, \bibinfo
  {author} {\bibfnamefont {X.}~\bibnamefont {Feng}}, \bibinfo {author}
  {\bibfnamefont {Z.}~\bibnamefont {Zhou}}, \bibinfo {author} {\bibfnamefont
  {L.}~\bibnamefont {An}}, \bibinfo {author} {\bibfnamefont {K.~T.}\
  \bibnamefont {Law}}, \bibinfo {author} {\bibfnamefont {N.}~\bibnamefont
  {Wang}},\ and\ \bibinfo {author} {\bibfnamefont {N.}~\bibnamefont {Lin}},\
  }\bibfield  {title} {\bibinfo {title} {Lattice reconstruction induced
  multiple ultra-flat bands in twisted bilayer wse2},\ }\bibfield  {journal}
  {\bibinfo  {journal} {Nature Communications}\ }\textbf {\bibinfo {volume}
  {12}},\ \href {https://doi.org/10.1038/s41467-021-25924-6}
  {10.1038/s41467-021-25924-6} (\bibinfo {year} {2021})\BibitemShut {NoStop}%
\bibitem [{\citenamefont {Bandurin}\ \emph {et~al.}(2022)\citenamefont
  {Bandurin}, \citenamefont {Principi}, \citenamefont {Phinney}, \citenamefont
  {Taniguchi}, \citenamefont {Watanabe},\ and\ \citenamefont
  {Jarillo-Herrero}}]{bandurin2022interlayer}%
  \BibitemOpen
  \bibfield  {author} {\bibinfo {author} {\bibfnamefont {D.}~\bibnamefont
  {Bandurin}}, \bibinfo {author} {\bibfnamefont {A.}~\bibnamefont {Principi}},
  \bibinfo {author} {\bibfnamefont {I.}~\bibnamefont {Phinney}}, \bibinfo
  {author} {\bibfnamefont {T.}~\bibnamefont {Taniguchi}}, \bibinfo {author}
  {\bibfnamefont {K.}~\bibnamefont {Watanabe}},\ and\ \bibinfo {author}
  {\bibfnamefont {P.}~\bibnamefont {Jarillo-Herrero}},\ }\bibfield  {title}
  {\bibinfo {title} {Interlayer electron-hole friction in tunable twisted
  bilayer graphene semimetal},\ }\href
  {https://doi.org/10.1103/PhysRevLett.129.206802} {\bibfield  {journal}
  {\bibinfo  {journal} {Physical Review Letters}\ }\textbf {\bibinfo {volume}
  {129}},\ \bibinfo {pages} {206802} (\bibinfo {year} {2022})}\BibitemShut
  {NoStop}%
\bibitem [{\citenamefont {Xu}\ \emph {et~al.}(2021)\citenamefont {Xu},
  \citenamefont {Al~Ezzi}, \citenamefont {Balakrishnan}, \citenamefont
  {Garcia-Ruiz}, \citenamefont {Tsim}, \citenamefont {Mullan}, \citenamefont
  {Barrier}, \citenamefont {Xin}, \citenamefont {Piot}, \citenamefont
  {Taniguchi}, \citenamefont {Watanabe}, \citenamefont {Carvalho},
  \citenamefont {Mishchenko}, \citenamefont {Geim}, \citenamefont {Fal'ko},
  \citenamefont {Adam}, \citenamefont {Neto}, \citenamefont {Novoselov},\ and\
  \citenamefont {Shi}}]{Xu2021}%
  \BibitemOpen
  \bibfield  {author} {\bibinfo {author} {\bibfnamefont {S.}~\bibnamefont
  {Xu}}, \bibinfo {author} {\bibfnamefont {M.~M.}\ \bibnamefont {Al~Ezzi}},
  \bibinfo {author} {\bibfnamefont {N.}~\bibnamefont {Balakrishnan}}, \bibinfo
  {author} {\bibfnamefont {A.}~\bibnamefont {Garcia-Ruiz}}, \bibinfo {author}
  {\bibfnamefont {B.}~\bibnamefont {Tsim}}, \bibinfo {author} {\bibfnamefont
  {C.}~\bibnamefont {Mullan}}, \bibinfo {author} {\bibfnamefont
  {J.}~\bibnamefont {Barrier}}, \bibinfo {author} {\bibfnamefont
  {N.}~\bibnamefont {Xin}}, \bibinfo {author} {\bibfnamefont {B.~A.}\
  \bibnamefont {Piot}}, \bibinfo {author} {\bibfnamefont {T.}~\bibnamefont
  {Taniguchi}}, \bibinfo {author} {\bibfnamefont {K.}~\bibnamefont {Watanabe}},
  \bibinfo {author} {\bibfnamefont {A.}~\bibnamefont {Carvalho}}, \bibinfo
  {author} {\bibfnamefont {A.}~\bibnamefont {Mishchenko}}, \bibinfo {author}
  {\bibfnamefont {A.~K.}\ \bibnamefont {Geim}}, \bibinfo {author}
  {\bibfnamefont {V.~I.}\ \bibnamefont {Fal'ko}}, \bibinfo {author}
  {\bibfnamefont {S.}~\bibnamefont {Adam}}, \bibinfo {author} {\bibfnamefont
  {A.~H.~C.}\ \bibnamefont {Neto}}, \bibinfo {author} {\bibfnamefont {K.~S.}\
  \bibnamefont {Novoselov}},\ and\ \bibinfo {author} {\bibfnamefont
  {Y.}~\bibnamefont {Shi}},\ }\bibfield  {title} {\bibinfo {title} {Tunable van
  hove singularities and correlated states in twisted monolayer--bilayer
  graphene},\ }\href {https://doi.org/10.1038/s41567-021-01172-9} {\bibfield
  {journal} {\bibinfo  {journal} {Nature Physics}\ }\textbf {\bibinfo {volume}
  {17}},\ \bibinfo {pages} {619} (\bibinfo {year} {2021})}\BibitemShut
  {NoStop}%
\bibitem [{\citenamefont {Chen}\ \emph {et~al.}(2021)\citenamefont {Chen},
  \citenamefont {He}, \citenamefont {Zhang}, \citenamefont {Hsieh},
  \citenamefont {Fei}, \citenamefont {Watanabe}, \citenamefont {Taniguchi},
  \citenamefont {Cobden}, \citenamefont {Xu}, \citenamefont {Dean} \emph
  {et~al.}}]{chen2021}%
  \BibitemOpen
  \bibfield  {author} {\bibinfo {author} {\bibfnamefont {S.}~\bibnamefont
  {Chen}}, \bibinfo {author} {\bibfnamefont {M.}~\bibnamefont {He}}, \bibinfo
  {author} {\bibfnamefont {Y.-H.}\ \bibnamefont {Zhang}}, \bibinfo {author}
  {\bibfnamefont {V.}~\bibnamefont {Hsieh}}, \bibinfo {author} {\bibfnamefont
  {Z.}~\bibnamefont {Fei}}, \bibinfo {author} {\bibfnamefont {K.}~\bibnamefont
  {Watanabe}}, \bibinfo {author} {\bibfnamefont {T.}~\bibnamefont {Taniguchi}},
  \bibinfo {author} {\bibfnamefont {D.~H.}\ \bibnamefont {Cobden}}, \bibinfo
  {author} {\bibfnamefont {X.}~\bibnamefont {Xu}}, \bibinfo {author}
  {\bibfnamefont {C.~R.}\ \bibnamefont {Dean}}, \emph {et~al.},\ }\bibfield
  {title} {\bibinfo {title} {Electrically tunable correlated and topological
  states in twisted monolayer--bilayer graphene},\ }\href
  {https://doi.org/10.1038/s41567-020-01062-6} {\bibfield  {journal} {\bibinfo
  {journal} {Nature Physics}\ }\textbf {\bibinfo {volume} {17}},\ \bibinfo
  {pages} {374} (\bibinfo {year} {2021})}\BibitemShut {NoStop}%
\bibitem [{\citenamefont {Woods}\ \emph {et~al.}(2021)\citenamefont {Woods},
  \citenamefont {Ares}, \citenamefont {Nevison-Andrews}, \citenamefont
  {Holwill}, \citenamefont {Fabregas}, \citenamefont {Guinea}, \citenamefont
  {Geim}, \citenamefont {Novoselov}, \citenamefont {Walet},\ and\ \citenamefont
  {Fumagalli}}]{woods2021}%
  \BibitemOpen
  \bibfield  {author} {\bibinfo {author} {\bibfnamefont {C.}~\bibnamefont
  {Woods}}, \bibinfo {author} {\bibfnamefont {P.}~\bibnamefont {Ares}},
  \bibinfo {author} {\bibfnamefont {H.}~\bibnamefont {Nevison-Andrews}},
  \bibinfo {author} {\bibfnamefont {M.}~\bibnamefont {Holwill}}, \bibinfo
  {author} {\bibfnamefont {R.}~\bibnamefont {Fabregas}}, \bibinfo {author}
  {\bibfnamefont {F.}~\bibnamefont {Guinea}}, \bibinfo {author} {\bibfnamefont
  {A.}~\bibnamefont {Geim}}, \bibinfo {author} {\bibfnamefont {K.}~\bibnamefont
  {Novoselov}}, \bibinfo {author} {\bibfnamefont {N.}~\bibnamefont {Walet}},\
  and\ \bibinfo {author} {\bibfnamefont {L.}~\bibnamefont {Fumagalli}},\
  }\bibfield  {title} {\bibinfo {title} {Charge-polarized interfacial
  superlattices in marginally twisted hexagonal boron nitride},\ }\href
  {https://doi.org/10.1038/s41467-020-20667-2} {\bibfield  {journal} {\bibinfo
  {journal} {Nature communications}\ }\textbf {\bibinfo {volume} {12}},\
  \bibinfo {pages} {1} (\bibinfo {year} {2021})}\BibitemShut {NoStop}%
\bibitem [{\citenamefont {Stern}\ \emph {et~al.}(2021)\citenamefont {Stern},
  \citenamefont {Waschitz}, \citenamefont {Cao}, \citenamefont {Nevo},
  \citenamefont {Watanabe}, \citenamefont {Taniguchi}, \citenamefont {Sela},
  \citenamefont {Urbakh}, \citenamefont {Hod},\ and\ \citenamefont
  {Shalom}}]{stern2021}%
  \BibitemOpen
  \bibfield  {author} {\bibinfo {author} {\bibfnamefont {M.~V.}\ \bibnamefont
  {Stern}}, \bibinfo {author} {\bibfnamefont {Y.}~\bibnamefont {Waschitz}},
  \bibinfo {author} {\bibfnamefont {W.}~\bibnamefont {Cao}}, \bibinfo {author}
  {\bibfnamefont {I.}~\bibnamefont {Nevo}}, \bibinfo {author} {\bibfnamefont
  {K.}~\bibnamefont {Watanabe}}, \bibinfo {author} {\bibfnamefont
  {T.}~\bibnamefont {Taniguchi}}, \bibinfo {author} {\bibfnamefont
  {E.}~\bibnamefont {Sela}}, \bibinfo {author} {\bibfnamefont {M.}~\bibnamefont
  {Urbakh}}, \bibinfo {author} {\bibfnamefont {O.}~\bibnamefont {Hod}},\ and\
  \bibinfo {author} {\bibfnamefont {M.~B.}\ \bibnamefont {Shalom}},\ }\bibfield
   {title} {\bibinfo {title} {Interfacial ferroelectricity by van der {W}aals
  sliding},\ }\href {https://doi.org/10.1126/science.abe8177} {\bibfield
  {journal} {\bibinfo  {journal} {Science}\ }\textbf {\bibinfo {volume}
  {372}},\ \bibinfo {pages} {1462} (\bibinfo {year} {2021})}\BibitemShut
  {NoStop}%
\bibitem [{\citenamefont {Yasuda}\ \emph {et~al.}(2021)\citenamefont {Yasuda},
  \citenamefont {Wang}, \citenamefont {Watanabe}, \citenamefont {Taniguchi},\
  and\ \citenamefont {Jarillo-Herrero}}]{yasuda2021}%
  \BibitemOpen
  \bibfield  {author} {\bibinfo {author} {\bibfnamefont {K.}~\bibnamefont
  {Yasuda}}, \bibinfo {author} {\bibfnamefont {X.}~\bibnamefont {Wang}},
  \bibinfo {author} {\bibfnamefont {K.}~\bibnamefont {Watanabe}}, \bibinfo
  {author} {\bibfnamefont {T.}~\bibnamefont {Taniguchi}},\ and\ \bibinfo
  {author} {\bibfnamefont {P.}~\bibnamefont {Jarillo-Herrero}},\ }\bibfield
  {title} {\bibinfo {title} {Stacking-engineered ferroelectricity in bilayer
  boron nitride},\ }\href {https://doi.org/10.1126/science.abd3230} {\bibfield
  {journal} {\bibinfo  {journal} {Science}\ }\textbf {\bibinfo {volume}
  {372}},\ \bibinfo {pages} {1458} (\bibinfo {year} {2021})}\BibitemShut
  {NoStop}%
\bibitem [{\citenamefont {Pan}\ \emph {et~al.}(2025)\citenamefont {Pan},
  \citenamefont {Li}, \citenamefont {Yang}, \citenamefont {Niu}, \citenamefont
  {Bian}, \citenamefont {Liu}, \citenamefont {Chen}, \citenamefont {Dong},
  \citenamefont {Wang}, \citenamefont {Zhou}, \citenamefont {Zhou},
  \citenamefont {Luo}, \citenamefont {Chu}, \citenamefont {Lin}, \citenamefont
  {Li},\ and\ \citenamefont {Liu}}]{Pan2025}%
  \BibitemOpen
  \bibfield  {author} {\bibinfo {author} {\bibfnamefont {E.}~\bibnamefont
  {Pan}}, \bibinfo {author} {\bibfnamefont {Z.}~\bibnamefont {Li}}, \bibinfo
  {author} {\bibfnamefont {F.}~\bibnamefont {Yang}}, \bibinfo {author}
  {\bibfnamefont {K.}~\bibnamefont {Niu}}, \bibinfo {author} {\bibfnamefont
  {R.}~\bibnamefont {Bian}}, \bibinfo {author} {\bibfnamefont {Q.}~\bibnamefont
  {Liu}}, \bibinfo {author} {\bibfnamefont {J.}~\bibnamefont {Chen}}, \bibinfo
  {author} {\bibfnamefont {B.}~\bibnamefont {Dong}}, \bibinfo {author}
  {\bibfnamefont {R.}~\bibnamefont {Wang}}, \bibinfo {author} {\bibfnamefont
  {T.}~\bibnamefont {Zhou}}, \bibinfo {author} {\bibfnamefont {A.}~\bibnamefont
  {Zhou}}, \bibinfo {author} {\bibfnamefont {X.}~\bibnamefont {Luo}}, \bibinfo
  {author} {\bibfnamefont {J.}~\bibnamefont {Chu}}, \bibinfo {author}
  {\bibfnamefont {J.}~\bibnamefont {Lin}}, \bibinfo {author} {\bibfnamefont
  {W.}~\bibnamefont {Li}},\ and\ \bibinfo {author} {\bibfnamefont
  {F.}~\bibnamefont {Liu}},\ }\bibfield  {title} {\bibinfo {title} {Observation
  and manipulation of two-dimensional topological polar texture confined in
  moiré interface},\ }\bibfield  {journal} {\bibinfo  {journal} {Nature
  Communications}\ }\textbf {\bibinfo {volume} {16}},\ \href
  {https://doi.org/10.1038/s41467-025-58105-w} {10.1038/s41467-025-58105-w}
  (\bibinfo {year} {2025})\BibitemShut {NoStop}%
\bibitem [{\citenamefont {Weston}\ \emph {et~al.}(2022)\citenamefont {Weston},
  \citenamefont {Castanon}, \citenamefont {Enaldiev}, \citenamefont {Ferreira},
  \citenamefont {Bhattacharjee}, \citenamefont {Xu}, \citenamefont
  {Corte-Le{\'{o}}n}, \citenamefont {Wu}, \citenamefont {Clark}, \citenamefont
  {Summerfield}, \citenamefont {Hashimoto}, \citenamefont {Gao}, \citenamefont
  {Wang}, \citenamefont {Hamer}, \citenamefont {Read}, \citenamefont
  {Fumagalli}, \citenamefont {Kretinin}, \citenamefont {Haigh}, \citenamefont
  {Kazakova}, \citenamefont {Geim}, \citenamefont {Fal'ko},\ and\ \citenamefont
  {Gorbachev}}]{Weston2022}%
  \BibitemOpen
  \bibfield  {author} {\bibinfo {author} {\bibfnamefont {A.}~\bibnamefont
  {Weston}}, \bibinfo {author} {\bibfnamefont {E.~G.}\ \bibnamefont
  {Castanon}}, \bibinfo {author} {\bibfnamefont {V.}~\bibnamefont {Enaldiev}},
  \bibinfo {author} {\bibfnamefont {F.}~\bibnamefont {Ferreira}}, \bibinfo
  {author} {\bibfnamefont {S.}~\bibnamefont {Bhattacharjee}}, \bibinfo {author}
  {\bibfnamefont {S.}~\bibnamefont {Xu}}, \bibinfo {author} {\bibfnamefont
  {H.}~\bibnamefont {Corte-Le{\'{o}}n}}, \bibinfo {author} {\bibfnamefont
  {Z.}~\bibnamefont {Wu}}, \bibinfo {author} {\bibfnamefont {N.}~\bibnamefont
  {Clark}}, \bibinfo {author} {\bibfnamefont {A.}~\bibnamefont {Summerfield}},
  \bibinfo {author} {\bibfnamefont {T.}~\bibnamefont {Hashimoto}}, \bibinfo
  {author} {\bibfnamefont {Y.}~\bibnamefont {Gao}}, \bibinfo {author}
  {\bibfnamefont {W.}~\bibnamefont {Wang}}, \bibinfo {author} {\bibfnamefont
  {M.}~\bibnamefont {Hamer}}, \bibinfo {author} {\bibfnamefont
  {H.}~\bibnamefont {Read}}, \bibinfo {author} {\bibfnamefont {L.}~\bibnamefont
  {Fumagalli}}, \bibinfo {author} {\bibfnamefont {A.~V.}\ \bibnamefont
  {Kretinin}}, \bibinfo {author} {\bibfnamefont {S.~J.}\ \bibnamefont {Haigh}},
  \bibinfo {author} {\bibfnamefont {O.}~\bibnamefont {Kazakova}}, \bibinfo
  {author} {\bibfnamefont {A.~K.}\ \bibnamefont {Geim}}, \bibinfo {author}
  {\bibfnamefont {V.~I.}\ \bibnamefont {Fal'ko}},\ and\ \bibinfo {author}
  {\bibfnamefont {R.}~\bibnamefont {Gorbachev}},\ }\bibfield  {title} {\bibinfo
  {title} {Interfacial ferroelectricity in marginally twisted 2d
  semiconductors},\ }\bibfield  {journal} {\bibinfo  {journal} {Nature
  Nanotechnology}\ }\href {https://doi.org/10.1038/s41565-022-01072-w}
  {10.1038/s41565-022-01072-w} (\bibinfo {year} {2022})\BibitemShut {NoStop}%
\bibitem [{\citenamefont {Wang}\ \emph {et~al.}(2022)\citenamefont {Wang},
  \citenamefont {Yasuda}, \citenamefont {Zhang}, \citenamefont {Liu},
  \citenamefont {Watanabe}, \citenamefont {Taniguchi}, \citenamefont {Hone},
  \citenamefont {Fu},\ and\ \citenamefont {Jarillo-Herrero}}]{Wang2022}%
  \BibitemOpen
  \bibfield  {author} {\bibinfo {author} {\bibfnamefont {X.}~\bibnamefont
  {Wang}}, \bibinfo {author} {\bibfnamefont {K.}~\bibnamefont {Yasuda}},
  \bibinfo {author} {\bibfnamefont {Y.}~\bibnamefont {Zhang}}, \bibinfo
  {author} {\bibfnamefont {S.}~\bibnamefont {Liu}}, \bibinfo {author}
  {\bibfnamefont {K.}~\bibnamefont {Watanabe}}, \bibinfo {author}
  {\bibfnamefont {T.}~\bibnamefont {Taniguchi}}, \bibinfo {author}
  {\bibfnamefont {J.}~\bibnamefont {Hone}}, \bibinfo {author} {\bibfnamefont
  {L.}~\bibnamefont {Fu}},\ and\ \bibinfo {author} {\bibfnamefont
  {P.}~\bibnamefont {Jarillo-Herrero}},\ }\bibfield  {title} {\bibinfo {title}
  {Interfacial ferroelectricity in rhombohedral-stacked bilayer transition
  metal dichalcogenides},\ }\href {https://doi.org/10.1038/s41565-021-01059-z}
  {\bibfield  {journal} {\bibinfo  {journal} {Nature Nanotechnology}\ }\textbf
  {\bibinfo {volume} {17}},\ \bibinfo {pages} {367} (\bibinfo {year}
  {2022})}\BibitemShut {NoStop}%
\bibitem [{\citenamefont {Vizner~Stern}\ \emph {et~al.}(2024)\citenamefont
  {Vizner~Stern}, \citenamefont {Salleh~Atri},\ and\ \citenamefont
  {Ben~Shalom}}]{ViznerStern2024}%
  \BibitemOpen
  \bibfield  {author} {\bibinfo {author} {\bibfnamefont {M.}~\bibnamefont
  {Vizner~Stern}}, \bibinfo {author} {\bibfnamefont {S.}~\bibnamefont
  {Salleh~Atri}},\ and\ \bibinfo {author} {\bibfnamefont {M.}~\bibnamefont
  {Ben~Shalom}},\ }\bibfield  {title} {\bibinfo {title} {Sliding van der waals
  polytypes},\ }\href {https://doi.org/10.1038/s42254-024-00781-6} {\bibfield
  {journal} {\bibinfo  {journal} {Nature Reviews Physics}\ }\textbf {\bibinfo
  {volume} {7}},\ \bibinfo {pages} {50} (\bibinfo {year} {2024})}\BibitemShut
  {NoStop}%
\bibitem [{\citenamefont {Deb}\ \emph {et~al.}(2022)\citenamefont {Deb},
  \citenamefont {Cao}, \citenamefont {Raab}, \citenamefont {Watanabe},
  \citenamefont {Taniguchi}, \citenamefont {Goldstein}, \citenamefont {Kronik},
  \citenamefont {Urbakh}, \citenamefont {Hod},\ and\ \citenamefont
  {Ben~Shalom}}]{Deb2022}%
  \BibitemOpen
  \bibfield  {author} {\bibinfo {author} {\bibfnamefont {S.}~\bibnamefont
  {Deb}}, \bibinfo {author} {\bibfnamefont {W.}~\bibnamefont {Cao}}, \bibinfo
  {author} {\bibfnamefont {N.}~\bibnamefont {Raab}}, \bibinfo {author}
  {\bibfnamefont {K.}~\bibnamefont {Watanabe}}, \bibinfo {author}
  {\bibfnamefont {T.}~\bibnamefont {Taniguchi}}, \bibinfo {author}
  {\bibfnamefont {M.}~\bibnamefont {Goldstein}}, \bibinfo {author}
  {\bibfnamefont {L.}~\bibnamefont {Kronik}}, \bibinfo {author} {\bibfnamefont
  {M.}~\bibnamefont {Urbakh}}, \bibinfo {author} {\bibfnamefont
  {O.}~\bibnamefont {Hod}},\ and\ \bibinfo {author} {\bibfnamefont
  {M.}~\bibnamefont {Ben~Shalom}},\ }\bibfield  {title} {\bibinfo {title}
  {Cumulative polarization in conductive interfacial ferroelectrics},\ }\href
  {https://doi.org/10.1038/s41586-022-05341-5} {\bibfield  {journal} {\bibinfo
  {journal} {Nature}\ }\textbf {\bibinfo {volume} {612}},\ \bibinfo {pages}
  {465} (\bibinfo {year} {2022})}\BibitemShut {NoStop}%
\bibitem [{\citenamefont {Van~Winkle}\ \emph {et~al.}(2024)\citenamefont
  {Van~Winkle}, \citenamefont {Dowlatshahi}, \citenamefont {Khaloo},
  \citenamefont {Iyer}, \citenamefont {Craig}, \citenamefont {Dhall},
  \citenamefont {Taniguchi}, \citenamefont {Watanabe},\ and\ \citenamefont
  {Bediako}}]{VanWinkle2024}%
  \BibitemOpen
  \bibfield  {author} {\bibinfo {author} {\bibfnamefont {M.}~\bibnamefont
  {Van~Winkle}}, \bibinfo {author} {\bibfnamefont {N.}~\bibnamefont
  {Dowlatshahi}}, \bibinfo {author} {\bibfnamefont {N.}~\bibnamefont {Khaloo}},
  \bibinfo {author} {\bibfnamefont {M.}~\bibnamefont {Iyer}}, \bibinfo {author}
  {\bibfnamefont {I.~M.}\ \bibnamefont {Craig}}, \bibinfo {author}
  {\bibfnamefont {R.}~\bibnamefont {Dhall}}, \bibinfo {author} {\bibfnamefont
  {T.}~\bibnamefont {Taniguchi}}, \bibinfo {author} {\bibfnamefont
  {K.}~\bibnamefont {Watanabe}},\ and\ \bibinfo {author} {\bibfnamefont
  {D.~K.}\ \bibnamefont {Bediako}},\ }\bibfield  {title} {\bibinfo {title}
  {Engineering interfacial polarization switching in van der waals
  multilayers},\ }\href {https://doi.org/10.1038/s41565-024-01642-0} {\bibfield
   {journal} {\bibinfo  {journal} {Nature Nanotechnology}\ }\textbf {\bibinfo
  {volume} {19}},\ \bibinfo {pages} {751} (\bibinfo {year} {2024})}\BibitemShut
  {NoStop}%
\bibitem [{\citenamefont {Ferreira}\ \emph {et~al.}(2022)\citenamefont
  {Ferreira}, \citenamefont {Enaldiev},\ and\ \citenamefont
  {Fal’ko}}]{Ferreira2022}%
  \BibitemOpen
  \bibfield  {author} {\bibinfo {author} {\bibfnamefont {F.}~\bibnamefont
  {Ferreira}}, \bibinfo {author} {\bibfnamefont {V.~V.}\ \bibnamefont
  {Enaldiev}},\ and\ \bibinfo {author} {\bibfnamefont {V.~I.}\ \bibnamefont
  {Fal’ko}},\ }\bibfield  {title} {\bibinfo {title} {Scaleability of
  dielectric susceptibility $\epsilon_{zz}$ with the number of layers and
  additivity of ferroelectric polarization in van der waals semiconductors},\
  }\href {https://doi.org/10.1103/physrevb.106.125408} {\bibfield  {journal}
  {\bibinfo  {journal} {Physical Review B}\ }\textbf {\bibinfo {volume}
  {106}},\ \bibinfo {pages} {125408} (\bibinfo {year} {2022})}\BibitemShut
  {NoStop}%
\bibitem [{\citenamefont {Enaldiev}\ \emph {et~al.}(2022)\citenamefont
  {Enaldiev}, \citenamefont {Ferreira},\ and\ \citenamefont
  {Fal'ko}}]{Enaldiev2022}%
  \BibitemOpen
  \bibfield  {author} {\bibinfo {author} {\bibfnamefont {V.~V.}\ \bibnamefont
  {Enaldiev}}, \bibinfo {author} {\bibfnamefont {F.}~\bibnamefont {Ferreira}},\
  and\ \bibinfo {author} {\bibfnamefont {V.~I.}\ \bibnamefont {Fal'ko}},\
  }\bibfield  {title} {\bibinfo {title} {A scalable network model for
  electrically tunable ferroelectric domain structure in twistronic bilayers of
  two-dimensional semiconductors},\ }\href
  {https://doi.org/10.1021/acs.nanolett.1c04210} {\bibfield  {journal}
  {\bibinfo  {journal} {Nano Letters}\ }\textbf {\bibinfo {volume} {22}},\
  \bibinfo {pages} {1534} (\bibinfo {year} {2022})}\BibitemShut {NoStop}%
\bibitem [{\citenamefont {Molino}\ \emph {et~al.}(2023)\citenamefont {Molino},
  \citenamefont {Aggarwal}, \citenamefont {Enaldiev}, \citenamefont
  {Plumadore}, \citenamefont {I.~Fal´ko},\ and\ \citenamefont
  {Luican‐Mayer}}]{Molino2023}%
  \BibitemOpen
  \bibfield  {author} {\bibinfo {author} {\bibfnamefont {L.}~\bibnamefont
  {Molino}}, \bibinfo {author} {\bibfnamefont {L.}~\bibnamefont {Aggarwal}},
  \bibinfo {author} {\bibfnamefont {V.}~\bibnamefont {Enaldiev}}, \bibinfo
  {author} {\bibfnamefont {R.}~\bibnamefont {Plumadore}}, \bibinfo {author}
  {\bibfnamefont {V.}~\bibnamefont {I.~Fal´ko}},\ and\ \bibinfo {author}
  {\bibfnamefont {A.}~\bibnamefont {Luican‐Mayer}},\ }\bibfield  {title}
  {\bibinfo {title} {Ferroelectric switching at symmetry‐broken interfaces by
  local control of dislocations networks},\ }\bibfield  {journal} {\bibinfo
  {journal} {Advanced Materials}\ }\textbf {\bibinfo {volume} {35}},\ \href
  {https://doi.org/10.1002/adma.202207816} {10.1002/adma.202207816} (\bibinfo
  {year} {2023})\BibitemShut {NoStop}%
\bibitem [{\citenamefont {Liang}\ \emph {et~al.}(2023)\citenamefont {Liang},
  \citenamefont {Yang}, \citenamefont {Xiao}, \citenamefont {Chen},
  \citenamefont {Dadap}, \citenamefont {Rottler},\ and\ \citenamefont
  {Ye}}]{Liang2023}%
  \BibitemOpen
  \bibfield  {author} {\bibinfo {author} {\bibfnamefont {J.}~\bibnamefont
  {Liang}}, \bibinfo {author} {\bibfnamefont {D.}~\bibnamefont {Yang}},
  \bibinfo {author} {\bibfnamefont {Y.}~\bibnamefont {Xiao}}, \bibinfo {author}
  {\bibfnamefont {S.}~\bibnamefont {Chen}}, \bibinfo {author} {\bibfnamefont
  {J.~I.}\ \bibnamefont {Dadap}}, \bibinfo {author} {\bibfnamefont
  {J.}~\bibnamefont {Rottler}},\ and\ \bibinfo {author} {\bibfnamefont
  {Z.}~\bibnamefont {Ye}},\ }\bibfield  {title} {\bibinfo {title} {Shear
  strain-induced two-dimensional slip avalanches in rhombohedral {MoS}$_2$},\
  }\href {https://doi.org/10.1021/acs.nanolett.3c01487} {\bibfield  {journal}
  {\bibinfo  {journal} {Nano Letters}\ }\textbf {\bibinfo {volume} {23}},\
  \bibinfo {pages} {7228} (\bibinfo {year} {2023})}\BibitemShut {NoStop}%
\bibitem [{\citenamefont {Ko}\ \emph {et~al.}(2023)\citenamefont {Ko},
  \citenamefont {Yuk}, \citenamefont {Engelke}, \citenamefont {Carr},
  \citenamefont {Kim}, \citenamefont {Park}, \citenamefont {Heo}, \citenamefont
  {Kim}, \citenamefont {Kim}, \citenamefont {Kim}, \citenamefont {Taniguchi},
  \citenamefont {Watanabe}, \citenamefont {Park}, \citenamefont {Kaxiras},
  \citenamefont {Yang}, \citenamefont {Kim},\ and\ \citenamefont
  {Yoo}}]{Ko2023}%
  \BibitemOpen
  \bibfield  {author} {\bibinfo {author} {\bibfnamefont {K.}~\bibnamefont
  {Ko}}, \bibinfo {author} {\bibfnamefont {A.}~\bibnamefont {Yuk}}, \bibinfo
  {author} {\bibfnamefont {R.}~\bibnamefont {Engelke}}, \bibinfo {author}
  {\bibfnamefont {S.}~\bibnamefont {Carr}}, \bibinfo {author} {\bibfnamefont
  {J.}~\bibnamefont {Kim}}, \bibinfo {author} {\bibfnamefont {D.}~\bibnamefont
  {Park}}, \bibinfo {author} {\bibfnamefont {H.}~\bibnamefont {Heo}}, \bibinfo
  {author} {\bibfnamefont {H.-M.}\ \bibnamefont {Kim}}, \bibinfo {author}
  {\bibfnamefont {S.-G.}\ \bibnamefont {Kim}}, \bibinfo {author} {\bibfnamefont
  {H.}~\bibnamefont {Kim}}, \bibinfo {author} {\bibfnamefont {T.}~\bibnamefont
  {Taniguchi}}, \bibinfo {author} {\bibfnamefont {K.}~\bibnamefont {Watanabe}},
  \bibinfo {author} {\bibfnamefont {H.}~\bibnamefont {Park}}, \bibinfo {author}
  {\bibfnamefont {E.}~\bibnamefont {Kaxiras}}, \bibinfo {author} {\bibfnamefont
  {S.~M.}\ \bibnamefont {Yang}}, \bibinfo {author} {\bibfnamefont
  {P.}~\bibnamefont {Kim}},\ and\ \bibinfo {author} {\bibfnamefont
  {H.}~\bibnamefont {Yoo}},\ }\bibfield  {title} {\bibinfo {title} {Operando
  electron microscopy investigation of polar domain dynamics in twisted van der
  waals homobilayers},\ }\href {https://doi.org/10.1038/s41563-023-01595-0}
  {\bibfield  {journal} {\bibinfo  {journal} {Nature Materials}\ }\textbf
  {\bibinfo {volume} {22}},\ \bibinfo {pages} {992} (\bibinfo {year}
  {2023})}\BibitemShut {NoStop}%
\bibitem [{\citenamefont {Bian}\ \emph {et~al.}(2024)\citenamefont {Bian},
  \citenamefont {He}, \citenamefont {Pan}, \citenamefont {Li}, \citenamefont
  {Cao}, \citenamefont {Meng}, \citenamefont {Chen}, \citenamefont {Liu},
  \citenamefont {Zhong}, \citenamefont {Li},\ and\ \citenamefont
  {Liu}}]{Bian2024}%
  \BibitemOpen
  \bibfield  {author} {\bibinfo {author} {\bibfnamefont {R.}~\bibnamefont
  {Bian}}, \bibinfo {author} {\bibfnamefont {R.}~\bibnamefont {He}}, \bibinfo
  {author} {\bibfnamefont {E.}~\bibnamefont {Pan}}, \bibinfo {author}
  {\bibfnamefont {Z.}~\bibnamefont {Li}}, \bibinfo {author} {\bibfnamefont
  {G.}~\bibnamefont {Cao}}, \bibinfo {author} {\bibfnamefont {P.}~\bibnamefont
  {Meng}}, \bibinfo {author} {\bibfnamefont {J.}~\bibnamefont {Chen}}, \bibinfo
  {author} {\bibfnamefont {Q.}~\bibnamefont {Liu}}, \bibinfo {author}
  {\bibfnamefont {Z.}~\bibnamefont {Zhong}}, \bibinfo {author} {\bibfnamefont
  {W.}~\bibnamefont {Li}},\ and\ \bibinfo {author} {\bibfnamefont
  {F.}~\bibnamefont {Liu}},\ }\bibfield  {title} {\bibinfo {title} {Developing
  fatigue-resistant ferroelectrics using interlayer sliding switching},\ }\href
  {https://doi.org/10.1126/science.ado1744} {\bibfield  {journal} {\bibinfo
  {journal} {Science}\ }\textbf {\bibinfo {volume} {385}},\ \bibinfo {pages}
  {57} (\bibinfo {year} {2024})}\BibitemShut {NoStop}%
\bibitem [{\citenamefont {Magorrian}\ \emph {et~al.}(2021)\citenamefont
  {Magorrian}, \citenamefont {Enaldiev}, \citenamefont {Z\'olyomi},
  \citenamefont {Ferreira}, \citenamefont {Fal’ko},\ and\ \citenamefont
  {Ruiz-Tijerina}}]{Magorrian2021}%
  \BibitemOpen
  \bibfield  {author} {\bibinfo {author} {\bibfnamefont {S.~J.}\ \bibnamefont
  {Magorrian}}, \bibinfo {author} {\bibfnamefont {V.~V.}\ \bibnamefont
  {Enaldiev}}, \bibinfo {author} {\bibfnamefont {V.}~\bibnamefont {Z\'olyomi}},
  \bibinfo {author} {\bibfnamefont {F.}~\bibnamefont {Ferreira}}, \bibinfo
  {author} {\bibfnamefont {V.~I.}\ \bibnamefont {Fal’ko}},\ and\ \bibinfo
  {author} {\bibfnamefont {D.~A.}\ \bibnamefont {Ruiz-Tijerina}},\ }\bibfield
  {title} {\bibinfo {title} {Multifaceted moiré superlattice physics in
  twisted {WS}e$_2$ bilayers},\ }\href
  {https://doi.org/10.1103/physrevb.104.125440} {\bibfield  {journal} {\bibinfo
   {journal} {Physical Review B}\ }\textbf {\bibinfo {volume} {104}},\ \bibinfo
  {pages} {125440} (\bibinfo {year} {2021})}\BibitemShut {NoStop}%
\bibitem [{\citenamefont {Zhao}\ \emph {et~al.}(2021)\citenamefont {Zhao},
  \citenamefont {Xiao},\ and\ \citenamefont {Yao}}]{Zhao2021}%
  \BibitemOpen
  \bibfield  {author} {\bibinfo {author} {\bibfnamefont {P.}~\bibnamefont
  {Zhao}}, \bibinfo {author} {\bibfnamefont {C.}~\bibnamefont {Xiao}},\ and\
  \bibinfo {author} {\bibfnamefont {W.}~\bibnamefont {Yao}},\ }\bibfield
  {title} {\bibinfo {title} {Universal superlattice potential for 2d materials
  from twisted interface inside h-bn substrate},\ }\bibfield  {journal}
  {\bibinfo  {journal} {npj 2D Materials and Applications}\ }\textbf {\bibinfo
  {volume} {5}},\ \href {https://doi.org/10.1038/s41699-021-00221-4}
  {10.1038/s41699-021-00221-4} (\bibinfo {year} {2021})\BibitemShut {NoStop}%
\bibitem [{\citenamefont {Ding}\ \emph {et~al.}(2024)\citenamefont {Ding},
  \citenamefont {Xiang}, \citenamefont {Zhou}, \citenamefont {Liu},
  \citenamefont {Chen}, \citenamefont {Fang}, \citenamefont {Wang},
  \citenamefont {Wu}, \citenamefont {Watanabe}, \citenamefont {Taniguchi},
  \citenamefont {Xin},\ and\ \citenamefont {Xu}}]{Ding2024}%
  \BibitemOpen
  \bibfield  {author} {\bibinfo {author} {\bibfnamefont {J.}~\bibnamefont
  {Ding}}, \bibinfo {author} {\bibfnamefont {H.}~\bibnamefont {Xiang}},
  \bibinfo {author} {\bibfnamefont {W.}~\bibnamefont {Zhou}}, \bibinfo {author}
  {\bibfnamefont {N.}~\bibnamefont {Liu}}, \bibinfo {author} {\bibfnamefont
  {Q.}~\bibnamefont {Chen}}, \bibinfo {author} {\bibfnamefont {X.}~\bibnamefont
  {Fang}}, \bibinfo {author} {\bibfnamefont {K.}~\bibnamefont {Wang}}, \bibinfo
  {author} {\bibfnamefont {L.}~\bibnamefont {Wu}}, \bibinfo {author}
  {\bibfnamefont {K.}~\bibnamefont {Watanabe}}, \bibinfo {author}
  {\bibfnamefont {T.}~\bibnamefont {Taniguchi}}, \bibinfo {author}
  {\bibfnamefont {N.}~\bibnamefont {Xin}},\ and\ \bibinfo {author}
  {\bibfnamefont {S.}~\bibnamefont {Xu}},\ }\bibfield  {title} {\bibinfo
  {title} {Engineering band structures of two-dimensional materials with remote
  moiré ferroelectricity},\ }\href
  {https://doi.org/10.1038/s41467-024-53440-w} {\bibfield  {journal} {\bibinfo
  {journal} {Nature Communications}\ }\textbf {\bibinfo {volume} {15}},\
  \bibinfo {pages} {9087} (\bibinfo {year} {2024})}\BibitemShut {NoStop}%
\bibitem [{\citenamefont {Wang}\ \emph {et~al.}(2025)\citenamefont {Wang},
  \citenamefont {Xu}, \citenamefont {Aronson}, \citenamefont {Bennett},
  \citenamefont {Paul}, \citenamefont {Crowley}, \citenamefont {Collignon},
  \citenamefont {Watanabe}, \citenamefont {Taniguchi}, \citenamefont {Ashoori},
  \citenamefont {Kaxiras}, \citenamefont {Zhang}, \citenamefont
  {Jarillo-Herrero},\ and\ \citenamefont {Yasuda}}]{Wang2025}%
  \BibitemOpen
  \bibfield  {author} {\bibinfo {author} {\bibfnamefont {X.}~\bibnamefont
  {Wang}}, \bibinfo {author} {\bibfnamefont {C.}~\bibnamefont {Xu}}, \bibinfo
  {author} {\bibfnamefont {S.}~\bibnamefont {Aronson}}, \bibinfo {author}
  {\bibfnamefont {D.}~\bibnamefont {Bennett}}, \bibinfo {author} {\bibfnamefont
  {N.}~\bibnamefont {Paul}}, \bibinfo {author} {\bibfnamefont {P.~J.~D.}\
  \bibnamefont {Crowley}}, \bibinfo {author} {\bibfnamefont {C.}~\bibnamefont
  {Collignon}}, \bibinfo {author} {\bibfnamefont {K.}~\bibnamefont {Watanabe}},
  \bibinfo {author} {\bibfnamefont {T.}~\bibnamefont {Taniguchi}}, \bibinfo
  {author} {\bibfnamefont {R.}~\bibnamefont {Ashoori}}, \bibinfo {author}
  {\bibfnamefont {E.}~\bibnamefont {Kaxiras}}, \bibinfo {author} {\bibfnamefont
  {Y.}~\bibnamefont {Zhang}}, \bibinfo {author} {\bibfnamefont
  {P.}~\bibnamefont {Jarillo-Herrero}},\ and\ \bibinfo {author} {\bibfnamefont
  {K.}~\bibnamefont {Yasuda}},\ }\bibfield  {title} {\bibinfo {title} {Moiré
  band structure engineering using a twisted boron nitride substrate},\
  }\bibfield  {journal} {\bibinfo  {journal} {Nature Communications}\ }\textbf
  {\bibinfo {volume} {16}},\ \href {https://doi.org/10.1038/s41467-024-55432-2}
  {10.1038/s41467-024-55432-2} (\bibinfo {year} {2025})\BibitemShut {NoStop}%
\bibitem [{\citenamefont {Kim}\ \emph {et~al.}(2023)\citenamefont {Kim},
  \citenamefont {Dominguez}, \citenamefont {Mayorga-Luna}, \citenamefont {Ye},
  \citenamefont {Embley}, \citenamefont {Tan}, \citenamefont {Ni},
  \citenamefont {Liu}, \citenamefont {Ford}, \citenamefont {Gao}, \citenamefont
  {Arash}, \citenamefont {Watanabe}, \citenamefont {Taniguchi}, \citenamefont
  {Kim}, \citenamefont {Shih}, \citenamefont {Lai}, \citenamefont {Yao},
  \citenamefont {Yang}, \citenamefont {Li},\ and\ \citenamefont
  {Miyahara}}]{Kim2023}%
  \BibitemOpen
  \bibfield  {author} {\bibinfo {author} {\bibfnamefont {D.~S.}\ \bibnamefont
  {Kim}}, \bibinfo {author} {\bibfnamefont {R.~C.}\ \bibnamefont {Dominguez}},
  \bibinfo {author} {\bibfnamefont {R.}~\bibnamefont {Mayorga-Luna}}, \bibinfo
  {author} {\bibfnamefont {D.}~\bibnamefont {Ye}}, \bibinfo {author}
  {\bibfnamefont {J.}~\bibnamefont {Embley}}, \bibinfo {author} {\bibfnamefont
  {T.}~\bibnamefont {Tan}}, \bibinfo {author} {\bibfnamefont {Y.}~\bibnamefont
  {Ni}}, \bibinfo {author} {\bibfnamefont {Z.}~\bibnamefont {Liu}}, \bibinfo
  {author} {\bibfnamefont {M.}~\bibnamefont {Ford}}, \bibinfo {author}
  {\bibfnamefont {F.~Y.}\ \bibnamefont {Gao}}, \bibinfo {author} {\bibfnamefont
  {S.}~\bibnamefont {Arash}}, \bibinfo {author} {\bibfnamefont
  {K.}~\bibnamefont {Watanabe}}, \bibinfo {author} {\bibfnamefont
  {T.}~\bibnamefont {Taniguchi}}, \bibinfo {author} {\bibfnamefont
  {S.}~\bibnamefont {Kim}}, \bibinfo {author} {\bibfnamefont {C.-K.}\
  \bibnamefont {Shih}}, \bibinfo {author} {\bibfnamefont {K.}~\bibnamefont
  {Lai}}, \bibinfo {author} {\bibfnamefont {W.}~\bibnamefont {Yao}}, \bibinfo
  {author} {\bibfnamefont {L.}~\bibnamefont {Yang}}, \bibinfo {author}
  {\bibfnamefont {X.}~\bibnamefont {Li}},\ and\ \bibinfo {author}
  {\bibfnamefont {Y.}~\bibnamefont {Miyahara}},\ }\bibfield  {title} {\bibinfo
  {title} {Electrostatic moiré potential from twisted hexagonal boron nitride
  layers},\ }\href {https://doi.org/10.1038/s41563-023-01637-7} {\bibfield
  {journal} {\bibinfo  {journal} {Nature Materials}\ }\textbf {\bibinfo
  {volume} {23}},\ \bibinfo {pages} {65} (\bibinfo {year} {2023})}\BibitemShut
  {NoStop}%
\bibitem [{\citenamefont {Kraut}\ and\ \citenamefont {von
  Baltz}(1979)}]{Kraut1979}%
  \BibitemOpen
  \bibfield  {author} {\bibinfo {author} {\bibfnamefont {W.}~\bibnamefont
  {Kraut}}\ and\ \bibinfo {author} {\bibfnamefont {R.}~\bibnamefont {von
  Baltz}},\ }\bibfield  {title} {\bibinfo {title} {Anomalous bulk photovoltaic
  effect in ferroelectrics: A quadratic response theory},\ }\href
  {https://doi.org/10.1103/physrevb.19.1548} {\bibfield  {journal} {\bibinfo
  {journal} {Physical Review B}\ }\textbf {\bibinfo {volume} {19}},\ \bibinfo
  {pages} {1548} (\bibinfo {year} {1979})}\BibitemShut {NoStop}%
\bibitem [{\citenamefont {von Baltz}\ and\ \citenamefont
  {Kraut}(1981)}]{Baltz1981}%
  \BibitemOpen
  \bibfield  {author} {\bibinfo {author} {\bibfnamefont {R.}~\bibnamefont {von
  Baltz}}\ and\ \bibinfo {author} {\bibfnamefont {W.}~\bibnamefont {Kraut}},\
  }\bibfield  {title} {\bibinfo {title} {Theory of the bulk photovoltaic effect
  in pure crystals},\ }\href {https://doi.org/10.1103/physrevb.23.5590}
  {\bibfield  {journal} {\bibinfo  {journal} {Physical Review B}\ }\textbf
  {\bibinfo {volume} {23}},\ \bibinfo {pages} {5590} (\bibinfo {year}
  {1981})}\BibitemShut {NoStop}%
\bibitem [{\citenamefont {Morimoto}\ and\ \citenamefont
  {Nagaosa}(2016)}]{morimoto2016topological}%
  \BibitemOpen
  \bibfield  {author} {\bibinfo {author} {\bibfnamefont {T.}~\bibnamefont
  {Morimoto}}\ and\ \bibinfo {author} {\bibfnamefont {N.}~\bibnamefont
  {Nagaosa}},\ }\bibfield  {title} {\bibinfo {title} {Topological nature of
  nonlinear optical effects in solids},\ }\href
  {https://doi.org/10.1126/sciadv.1501524} {\bibfield  {journal} {\bibinfo
  {journal} {Science advances}\ }\textbf {\bibinfo {volume} {2}},\ \bibinfo
  {pages} {e1501524} (\bibinfo {year} {2016})}\BibitemShut {NoStop}%
\bibitem [{\citenamefont {Ma}\ \emph {et~al.}(2023)\citenamefont {Ma},
  \citenamefont {Krishna~Kumar}, \citenamefont {Xu}, \citenamefont {Koppens},\
  and\ \citenamefont {Song}}]{Ma2023}%
  \BibitemOpen
  \bibfield  {author} {\bibinfo {author} {\bibfnamefont {Q.}~\bibnamefont
  {Ma}}, \bibinfo {author} {\bibfnamefont {R.}~\bibnamefont {Krishna~Kumar}},
  \bibinfo {author} {\bibfnamefont {S.-Y.}\ \bibnamefont {Xu}}, \bibinfo
  {author} {\bibfnamefont {F.~H.~L.}\ \bibnamefont {Koppens}},\ and\ \bibinfo
  {author} {\bibfnamefont {J.~C.~W.}\ \bibnamefont {Song}},\ }\bibfield
  {title} {\bibinfo {title} {Photocurrent as a multiphysics diagnostic of
  quantum materials},\ }\href {https://doi.org/10.1038/s42254-022-00551-2}
  {\bibfield  {journal} {\bibinfo  {journal} {Nature Reviews Physics}\ }\textbf
  {\bibinfo {volume} {5}},\ \bibinfo {pages} {170} (\bibinfo {year}
  {2023})}\BibitemShut {NoStop}%
\bibitem [{\citenamefont {Castro~Neto}\ \emph {et~al.}(2009)\citenamefont
  {Castro~Neto}, \citenamefont {Guinea}, \citenamefont {Peres}, \citenamefont
  {Novoselov},\ and\ \citenamefont {Geim}}]{CastroNeto2009}%
  \BibitemOpen
  \bibfield  {author} {\bibinfo {author} {\bibfnamefont {A.~H.}\ \bibnamefont
  {Castro~Neto}}, \bibinfo {author} {\bibfnamefont {F.}~\bibnamefont {Guinea}},
  \bibinfo {author} {\bibfnamefont {N.~M.~R.}\ \bibnamefont {Peres}}, \bibinfo
  {author} {\bibfnamefont {K.~S.}\ \bibnamefont {Novoselov}},\ and\ \bibinfo
  {author} {\bibfnamefont {A.~K.}\ \bibnamefont {Geim}},\ }\bibfield  {title}
  {\bibinfo {title} {The electronic properties of graphene},\ }\href
  {https://doi.org/10.1103/revmodphys.81.109} {\bibfield  {journal} {\bibinfo
  {journal} {Reviews of Modern Physics}\ }\textbf {\bibinfo {volume} {81}},\
  \bibinfo {pages} {109} (\bibinfo {year} {2009})}\BibitemShut {NoStop}%
\bibitem [{fn1()}]{fn1}%
  \BibitemOpen
  \href@noop {} {\ }\bibinfo {note} {We use a representation with the same
  Hamiltonian for $\pm{\rm K}$-valleys of graphene}\BibitemShut {NoStop}%
\bibitem [{\citenamefont {Ortix}\ \emph {et~al.}(2012)\citenamefont {Ortix},
  \citenamefont {Yang},\ and\ \citenamefont {van~den Brink}}]{Ortix2012}%
  \BibitemOpen
  \bibfield  {author} {\bibinfo {author} {\bibfnamefont {C.}~\bibnamefont
  {Ortix}}, \bibinfo {author} {\bibfnamefont {L.}~\bibnamefont {Yang}},\ and\
  \bibinfo {author} {\bibfnamefont {J.}~\bibnamefont {van~den Brink}},\
  }\bibfield  {title} {\bibinfo {title} {Graphene on incommensurate substrates:
  Trigonal warping and emerging dirac cone replicas with halved group
  velocity},\ }\href {https://doi.org/10.1103/physrevb.86.081405} {\bibfield
  {journal} {\bibinfo  {journal} {Physical Review B}\ }\textbf {\bibinfo
  {volume} {86}},\ \bibinfo {pages} {081405} (\bibinfo {year}
  {2012})}\BibitemShut {NoStop}%
\bibitem [{\citenamefont {Wallbank}\ \emph {et~al.}(2013)\citenamefont
  {Wallbank}, \citenamefont {Patel}, \citenamefont {Mucha-Kruczyński},
  \citenamefont {Geim},\ and\ \citenamefont {Fal’ko}}]{Wallbank2013}%
  \BibitemOpen
  \bibfield  {author} {\bibinfo {author} {\bibfnamefont {J.~R.}\ \bibnamefont
  {Wallbank}}, \bibinfo {author} {\bibfnamefont {A.~A.}\ \bibnamefont {Patel}},
  \bibinfo {author} {\bibfnamefont {M.}~\bibnamefont {Mucha-Kruczyński}},
  \bibinfo {author} {\bibfnamefont {A.~K.}\ \bibnamefont {Geim}},\ and\
  \bibinfo {author} {\bibfnamefont {V.~I.}\ \bibnamefont {Fal’ko}},\
  }\bibfield  {title} {\bibinfo {title} {Generic miniband structure of graphene
  on a hexagonal substrate},\ }\href
  {https://doi.org/10.1103/physrevb.87.245408} {\bibfield  {journal} {\bibinfo
  {journal} {Physical Review B}\ }\textbf {\bibinfo {volume} {87}},\ \bibinfo
  {pages} {245408} (\bibinfo {year} {2013})}\BibitemShut {NoStop}%
\bibitem [{\citenamefont {Park}\ \emph
  {et~al.}(2008{\natexlab{a}})\citenamefont {Park}, \citenamefont {Yang},
  \citenamefont {Son}, \citenamefont {Cohen},\ and\ \citenamefont
  {Louie}}]{Park2008a}%
  \BibitemOpen
  \bibfield  {author} {\bibinfo {author} {\bibfnamefont {C.-H.}\ \bibnamefont
  {Park}}, \bibinfo {author} {\bibfnamefont {L.}~\bibnamefont {Yang}}, \bibinfo
  {author} {\bibfnamefont {Y.-W.}\ \bibnamefont {Son}}, \bibinfo {author}
  {\bibfnamefont {M.~L.}\ \bibnamefont {Cohen}},\ and\ \bibinfo {author}
  {\bibfnamefont {S.~G.}\ \bibnamefont {Louie}},\ }\bibfield  {title} {\bibinfo
  {title} {Anisotropic behaviours of massless dirac fermions in graphene under
  periodic potentials},\ }\href {https://doi.org/10.1038/nphys890} {\bibfield
  {journal} {\bibinfo  {journal} {Nature Physics}\ }\textbf {\bibinfo {volume}
  {4}},\ \bibinfo {pages} {213} (\bibinfo {year}
  {2008}{\natexlab{a}})}\BibitemShut {NoStop}%
\bibitem [{\citenamefont {Park}\ \emph
  {et~al.}(2008{\natexlab{b}})\citenamefont {Park}, \citenamefont {Yang},
  \citenamefont {Son}, \citenamefont {Cohen},\ and\ \citenamefont
  {Louie}}]{Park2008}%
  \BibitemOpen
  \bibfield  {author} {\bibinfo {author} {\bibfnamefont {C.-H.}\ \bibnamefont
  {Park}}, \bibinfo {author} {\bibfnamefont {L.}~\bibnamefont {Yang}}, \bibinfo
  {author} {\bibfnamefont {Y.-W.}\ \bibnamefont {Son}}, \bibinfo {author}
  {\bibfnamefont {M.~L.}\ \bibnamefont {Cohen}},\ and\ \bibinfo {author}
  {\bibfnamefont {S.~G.}\ \bibnamefont {Louie}},\ }\bibfield  {title} {\bibinfo
  {title} {New generation of massless dirac fermions in graphene under external
  periodic potentials},\ }\href
  {https://doi.org/10.1103/physrevlett.101.126804} {\bibfield  {journal}
  {\bibinfo  {journal} {Physical Review Letters}\ }\textbf {\bibinfo {volume}
  {101}},\ \bibinfo {pages} {126804} (\bibinfo {year}
  {2008}{\natexlab{b}})}\BibitemShut {NoStop}%
\bibitem [{\citenamefont {Cheianov}\ and\ \citenamefont
  {Fal’ko}(2006)}]{Cheianov2006}%
  \BibitemOpen
  \bibfield  {author} {\bibinfo {author} {\bibfnamefont {V.~V.}\ \bibnamefont
  {Cheianov}}\ and\ \bibinfo {author} {\bibfnamefont {V.~I.}\ \bibnamefont
  {Fal’ko}},\ }\bibfield  {title} {\bibinfo {title} {Selective transmission
  of dirac electrons and ballistic magnetoresistance of n-p junctions in
  graphene},\ }\href {https://doi.org/10.1103/physrevb.74.041403} {\bibfield
  {journal} {\bibinfo  {journal} {Physical Review B}\ }\textbf {\bibinfo
  {volume} {74}},\ \bibinfo {pages} {041403} (\bibinfo {year}
  {2006})}\BibitemShut {NoStop}%
\bibitem [{\citenamefont {Lukose}\ \emph {et~al.}(2007)\citenamefont {Lukose},
  \citenamefont {Shankar},\ and\ \citenamefont {Baskaran}}]{Lukose2007}%
  \BibitemOpen
  \bibfield  {author} {\bibinfo {author} {\bibfnamefont {V.}~\bibnamefont
  {Lukose}}, \bibinfo {author} {\bibfnamefont {R.}~\bibnamefont {Shankar}},\
  and\ \bibinfo {author} {\bibfnamefont {G.}~\bibnamefont {Baskaran}},\
  }\bibfield  {title} {\bibinfo {title} {Novel electric field effects on landau
  levels in graphene},\ }\href {https://doi.org/10.1103/physrevlett.98.116802}
  {\bibfield  {journal} {\bibinfo  {journal} {Physical Review Letters}\
  }\textbf {\bibinfo {volume} {98}},\ \bibinfo {pages} {116802} (\bibinfo
  {year} {2007})}\BibitemShut {NoStop}%
\bibitem [{\citenamefont {Alisultanov}(2014)}]{alisultanov2014landau}%
  \BibitemOpen
  \bibfield  {author} {\bibinfo {author} {\bibfnamefont {Z.}~\bibnamefont
  {Alisultanov}},\ }\bibfield  {title} {\bibinfo {title} {Landau levels in
  graphene in crossed magnetic and electric fields: Quasi-classical approach},\
  }\href {https://doi.org/10.1016/j.physb.2013.12.033} {\bibfield  {journal}
  {\bibinfo  {journal} {Physica B: Condensed Matter}\ }\textbf {\bibinfo
  {volume} {438}},\ \bibinfo {pages} {41} (\bibinfo {year} {2014})}\BibitemShut
  {NoStop}%
\bibitem [{fn2()}]{fn2}%
  \BibitemOpen
  \href@noop {} {\ }\bibinfo {note} {The Hamiltonian in vicinity of the other
  non-equivalent points of miniBZ,
  $\bm{\mu}_{1,2}=\hat{R}_{2\pi/3}^{\pm1}\bm{\mu}_3$, can be obtained using
  $H(\bm{\mu}_3+\bm{q})=H(\bm{\mu}_{1,2}+\hat{R}^{\pm
  1}_{2\pi/3}\bm{q})$}\BibitemShut {NoStop}%
\bibitem [{\citenamefont {Yankowitz}\ \emph {et~al.}(2012)\citenamefont
  {Yankowitz}, \citenamefont {Xue}, \citenamefont {Cormode}, \citenamefont
  {Sanchez-Yamagishi}, \citenamefont {Watanabe}, \citenamefont {Taniguchi},
  \citenamefont {Jarillo-Herrero}, \citenamefont {Jacquod},\ and\ \citenamefont
  {LeRoy}}]{Yankowitz2012}%
  \BibitemOpen
  \bibfield  {author} {\bibinfo {author} {\bibfnamefont {M.}~\bibnamefont
  {Yankowitz}}, \bibinfo {author} {\bibfnamefont {J.}~\bibnamefont {Xue}},
  \bibinfo {author} {\bibfnamefont {D.}~\bibnamefont {Cormode}}, \bibinfo
  {author} {\bibfnamefont {J.~D.}\ \bibnamefont {Sanchez-Yamagishi}}, \bibinfo
  {author} {\bibfnamefont {K.}~\bibnamefont {Watanabe}}, \bibinfo {author}
  {\bibfnamefont {T.}~\bibnamefont {Taniguchi}}, \bibinfo {author}
  {\bibfnamefont {P.}~\bibnamefont {Jarillo-Herrero}}, \bibinfo {author}
  {\bibfnamefont {P.}~\bibnamefont {Jacquod}},\ and\ \bibinfo {author}
  {\bibfnamefont {B.~J.}\ \bibnamefont {LeRoy}},\ }\bibfield  {title} {\bibinfo
  {title} {Emergence of superlattice dirac points in graphene on hexagonal
  boron nitride},\ }\href {https://doi.org/10.1038/nphys2272} {\bibfield
  {journal} {\bibinfo  {journal} {Nature Physics}\ }\textbf {\bibinfo {volume}
  {8}},\ \bibinfo {pages} {382} (\bibinfo {year} {2012})}\BibitemShut {NoStop}%
\bibitem [{\citenamefont {Callaway}(1964)}]{Callaway1964}%
  \BibitemOpen
  \bibfield  {author} {\bibinfo {author} {\bibfnamefont {J.}~\bibnamefont
  {Callaway}},\ }\href@noop {} {\emph {\bibinfo {title} {Energy band theory}}}\
  (\bibinfo  {publisher} {Academic Press},\ \bibinfo {address} {New York and
  London},\ \bibinfo {year} {1964})\ \bibinfo {note} {chapter 4,
  \S9}\BibitemShut {NoStop}%
\bibitem [{\citenamefont {Nair}\ \emph {et~al.}(2008)\citenamefont {Nair},
  \citenamefont {Blake}, \citenamefont {Grigorenko}, \citenamefont {Novoselov},
  \citenamefont {Booth}, \citenamefont {Stauber}, \citenamefont {Peres},\ and\
  \citenamefont {Geim}}]{Nair2008}%
  \BibitemOpen
  \bibfield  {author} {\bibinfo {author} {\bibfnamefont {R.~R.}\ \bibnamefont
  {Nair}}, \bibinfo {author} {\bibfnamefont {P.}~\bibnamefont {Blake}},
  \bibinfo {author} {\bibfnamefont {A.~N.}\ \bibnamefont {Grigorenko}},
  \bibinfo {author} {\bibfnamefont {K.~S.}\ \bibnamefont {Novoselov}}, \bibinfo
  {author} {\bibfnamefont {T.~J.}\ \bibnamefont {Booth}}, \bibinfo {author}
  {\bibfnamefont {T.}~\bibnamefont {Stauber}}, \bibinfo {author} {\bibfnamefont
  {N.~M.~R.}\ \bibnamefont {Peres}},\ and\ \bibinfo {author} {\bibfnamefont
  {A.~K.}\ \bibnamefont {Geim}},\ }\bibfield  {title} {\bibinfo {title} {Fine
  structure constant defines visual transparency of graphene},\ }\href
  {https://doi.org/10.1126/science.1156965} {\bibfield  {journal} {\bibinfo
  {journal} {Science}\ }\textbf {\bibinfo {volume} {320}},\ \bibinfo {pages}
  {1308} (\bibinfo {year} {2008})}\BibitemShut {NoStop}%
\bibitem [{\citenamefont {Ju}\ \emph {et~al.}(2011)\citenamefont {Ju},
  \citenamefont {Geng}, \citenamefont {Horng}, \citenamefont {Girit},
  \citenamefont {Martin}, \citenamefont {Hao}, \citenamefont {Bechtel},
  \citenamefont {Liang}, \citenamefont {Zettl}, \citenamefont {Shen},\ and\
  \citenamefont {Wang}}]{Ju2011}%
  \BibitemOpen
  \bibfield  {author} {\bibinfo {author} {\bibfnamefont {L.}~\bibnamefont
  {Ju}}, \bibinfo {author} {\bibfnamefont {B.}~\bibnamefont {Geng}}, \bibinfo
  {author} {\bibfnamefont {J.}~\bibnamefont {Horng}}, \bibinfo {author}
  {\bibfnamefont {C.}~\bibnamefont {Girit}}, \bibinfo {author} {\bibfnamefont
  {M.}~\bibnamefont {Martin}}, \bibinfo {author} {\bibfnamefont
  {Z.}~\bibnamefont {Hao}}, \bibinfo {author} {\bibfnamefont {H.~A.}\
  \bibnamefont {Bechtel}}, \bibinfo {author} {\bibfnamefont {X.}~\bibnamefont
  {Liang}}, \bibinfo {author} {\bibfnamefont {A.}~\bibnamefont {Zettl}},
  \bibinfo {author} {\bibfnamefont {Y.~R.}\ \bibnamefont {Shen}},\ and\
  \bibinfo {author} {\bibfnamefont {F.}~\bibnamefont {Wang}},\ }\bibfield
  {title} {\bibinfo {title} {Graphene plasmonics for tunable terahertz
  metamaterials},\ }\href {https://doi.org/10.1038/nnano.2011.146} {\bibfield
  {journal} {\bibinfo  {journal} {Nature Nanotechnology}\ }\textbf {\bibinfo
  {volume} {6}},\ \bibinfo {pages} {630} (\bibinfo {year} {2011})}\BibitemShut
  {NoStop}%
\bibitem [{fn3()}]{fn3}%
  \BibitemOpen
  \href@noop {} {\ }\bibinfo {note}
  {$\hat{T}\psi_{n,\bm{k}}(\bm{r})=e^{i\alpha_{n}(\bm{k})}\psi_{n,-\bm{k}}(\bm{r})$,
  where $\alpha_{n}(\bm{k})$ is a gauge phase.}\BibitemShut {Stop}%
\bibitem [{\citenamefont {Golub}\ \emph {et~al.}(2011)\citenamefont {Golub},
  \citenamefont {Tarasenko}, \citenamefont {Entin},\ and\ \citenamefont
  {Magarill}}]{Golub2011}%
  \BibitemOpen
  \bibfield  {author} {\bibinfo {author} {\bibfnamefont {L.}~\bibnamefont
  {Golub}}, \bibinfo {author} {\bibfnamefont {S.}~\bibnamefont {Tarasenko}},
  \bibinfo {author} {\bibfnamefont {M.}~\bibnamefont {Entin}},\ and\ \bibinfo
  {author} {\bibfnamefont {L.}~\bibnamefont {Magarill}},\ }\bibfield  {title}
  {\bibinfo {title} {Valley separation in graphene by polarized light},\ }\href
  {https://doi.org/10.1103/physrevb.84.195408} {\bibfield  {journal} {\bibinfo
  {journal} {Physical Review B}\ }\textbf {\bibinfo {volume} {84}},\ \bibinfo
  {pages} {195408} (\bibinfo {year} {2011})}\BibitemShut {NoStop}%
\bibitem [{fn4()}]{fn4}%
  \BibitemOpen
  \href@noop {} {\ }\bibinfo {note} {In the used representation of Hamiltonian
  \eqref{Eq:Ham} the trigonal warping term is proportional to
  $\mp\left[\sigma_x(\hat{p}_x^2-\hat{p}_y^2)-2\sigma_y\hat{p}_x\hat{p}_y\right]$
  for $\pm {\rm K}$-valley, leading to opposite corrections to the velocity
  operator and zero total injection photocurrent.}\BibitemShut {Stop}%
\bibitem [{fn5()}]{fn5}%
  \BibitemOpen
  \href@noop {} {\ }\bibinfo {note} {For $\sigma_{ijk}$, relating
  $j_i=\sigma_{ijk}E_jE^*_k$, there is the only independent in-plane component
  as $\sigma_{xxx}=-\sigma_{xyy}=-\sigma_{yxy}=-\sigma_{yyx}$.}\BibitemShut
  {Stop}%
\bibitem [{\citenamefont {Ferreira}\ \emph {et~al.}(2021)\citenamefont
  {Ferreira}, \citenamefont {Enaldiev}, \citenamefont {Fal'ko},\ and\
  \citenamefont {Magorrian}}]{Ferreira2021}%
  \BibitemOpen
  \bibfield  {author} {\bibinfo {author} {\bibfnamefont {F.}~\bibnamefont
  {Ferreira}}, \bibinfo {author} {\bibfnamefont {V.~V.}\ \bibnamefont
  {Enaldiev}}, \bibinfo {author} {\bibfnamefont {V.~I.}\ \bibnamefont
  {Fal'ko}},\ and\ \bibinfo {author} {\bibfnamefont {S.~J.}\ \bibnamefont
  {Magorrian}},\ }\bibfield  {title} {\bibinfo {title} {Weak ferroelectric
  charge transfer in layer-asymmetric bilayers of 2{D} semiconductors},\ }\href
  {https://doi.org/10.1038/s41598-021-92710-1} {\bibfield  {journal} {\bibinfo
  {journal} {Scientific Reports}\ }\textbf {\bibinfo {volume} {11}},\ \bibinfo
  {pages} {13422} (\bibinfo {year} {2021})}\BibitemShut {NoStop}%
\bibitem [{\citenamefont {Enaldiev}\ \emph {et~al.}(2020)\citenamefont
  {Enaldiev}, \citenamefont {Z\'olyomi}, \citenamefont {Yelgel}, \citenamefont
  {Magorrian},\ and\ \citenamefont {Fal'ko}}]{Enaldiev_PRL}%
  \BibitemOpen
  \bibfield  {author} {\bibinfo {author} {\bibfnamefont {V.~V.}\ \bibnamefont
  {Enaldiev}}, \bibinfo {author} {\bibfnamefont {V.}~\bibnamefont {Z\'olyomi}},
  \bibinfo {author} {\bibfnamefont {C.}~\bibnamefont {Yelgel}}, \bibinfo
  {author} {\bibfnamefont {S.~J.}\ \bibnamefont {Magorrian}},\ and\ \bibinfo
  {author} {\bibfnamefont {V.~I.}\ \bibnamefont {Fal'ko}},\ }\bibfield  {title}
  {\bibinfo {title} {Stacking domains and dislocation networks in marginally
  twisted bilayers of transition metal dichalcogenides},\ }\href
  {https://doi.org/10.1103/PhysRevLett.124.206101} {\bibfield  {journal}
  {\bibinfo  {journal} {Phys. Rev. Lett.}\ }\textbf {\bibinfo {volume} {124}},\
  \bibinfo {pages} {206101} (\bibinfo {year} {2020})}\BibitemShut {NoStop}%
\bibitem [{\citenamefont {Duerloo}\ \emph {et~al.}(2012)\citenamefont
  {Duerloo}, \citenamefont {Ong},\ and\ \citenamefont {Reed}}]{Duerloo2012}%
  \BibitemOpen
  \bibfield  {author} {\bibinfo {author} {\bibfnamefont {K.-A.~N.}\
  \bibnamefont {Duerloo}}, \bibinfo {author} {\bibfnamefont {M.~T.}\
  \bibnamefont {Ong}},\ and\ \bibinfo {author} {\bibfnamefont {E.~J.}\
  \bibnamefont {Reed}},\ }\bibfield  {title} {\bibinfo {title} {Intrinsic
  piezoelectricity in two-dimensional materials},\ }\href
  {https://doi.org/10.1021/jz3012436} {\bibfield  {journal} {\bibinfo
  {journal} {The Journal of Physical Chemistry Letters}\ }\textbf {\bibinfo
  {volume} {3}},\ \bibinfo {pages} {2871} (\bibinfo {year} {2012})}\BibitemShut
  {NoStop}%
\bibitem [{\citenamefont {Zhu}\ \emph {et~al.}(2014)\citenamefont {Zhu},
  \citenamefont {Wang}, \citenamefont {Xiao}, \citenamefont {Liu},
  \citenamefont {Xiong}, \citenamefont {Wong}, \citenamefont {Ye},
  \citenamefont {Ye}, \citenamefont {Yin},\ and\ \citenamefont
  {Zhang}}]{Zhu2014}%
  \BibitemOpen
  \bibfield  {author} {\bibinfo {author} {\bibfnamefont {H.}~\bibnamefont
  {Zhu}}, \bibinfo {author} {\bibfnamefont {Y.}~\bibnamefont {Wang}}, \bibinfo
  {author} {\bibfnamefont {J.}~\bibnamefont {Xiao}}, \bibinfo {author}
  {\bibfnamefont {M.}~\bibnamefont {Liu}}, \bibinfo {author} {\bibfnamefont
  {S.}~\bibnamefont {Xiong}}, \bibinfo {author} {\bibfnamefont {Z.~J.}\
  \bibnamefont {Wong}}, \bibinfo {author} {\bibfnamefont {Z.}~\bibnamefont
  {Ye}}, \bibinfo {author} {\bibfnamefont {Y.}~\bibnamefont {Ye}}, \bibinfo
  {author} {\bibfnamefont {X.}~\bibnamefont {Yin}},\ and\ \bibinfo {author}
  {\bibfnamefont {X.}~\bibnamefont {Zhang}},\ }\bibfield  {title} {\bibinfo
  {title} {Observation of piezoelectricity in free-standing monolayer mos2},\
  }\href {https://doi.org/10.1038/nnano.2014.309} {\bibfield  {journal}
  {\bibinfo  {journal} {Nature Nanotechnology}\ }\textbf {\bibinfo {volume}
  {10}},\ \bibinfo {pages} {151} (\bibinfo {year} {2014})}\BibitemShut
  {NoStop}%
\end{thebibliography}%
\renewcommand{\thesection}{S\arabic{section}}
\renewcommand{\thetable}{S\arabic{table}}
\renewcommand{\thefigure}{S\arabic{figure}}
\renewcommand{\theequation}{S\arabic{equation}}
\newcommand{\ch}{{\rm ch}}
\newcommand{\sh}{{\rm sh}}

\begin{widetext}

\begin{center}
	{\bf Supplementary Material for manuscript ''Resonant absorption and linear photovoltaic effect in ferroelectric moir\'e heterostructures''}
\end{center}

\section{Electrostatic moir\'e potential in graphene from polar domains in ferroelelctric twisted bilayer hexagonal borone nitride (hBN) or transition metal dichalcogenide (TMD)}

In Fig. \ref{fig:electrostatics} we display sketch of the electrostatic model. Lattice relaxation in parallel twisted hBN and TMD bilayers results in formation of polar domains characterized by alternating (up and down) out-of-plane ferroelectric polarization. To describe the in-plane distribution of the polarization in the twisted ferroelectric bilayers, with misalignment angle $\theta$  between layers, we use the following formula \cite{Ferreira2021,Magorrian2021}:
\begin{equation} \label{eqS:polarization}
    P(\bm{r})=\frac{2P_0}{3\sqrt{3}}\sum_{n=1,2,3}\sin\left(\bm{G}_n\bm{r}_0(\bm{r})\right),
\end{equation}
where $P_0$ is the magnitude of ferroelectric polarization per unit area inside domains, $\bm{r}_0(\bm{r}) = \theta \hat{z}\times\bm{r}+\bm{u}^t-\bm{u}^b$ is a local interlayer lateral offset characterizing local stacking of the layers in moir\'e superlattice accounting of relaxation of atomic positions described by displacement fields, $\bm{u}^{t/b}$, in t/b layers, $\bm{G}_{1,2,3}$ are a triad of the shortest reciprocal vectors of a single layer related by 120$^\circ$-rotation ($P_0\approx2.25 \times10^{-12}$\,C/m for hBN bilayers \cite{yasuda2021}, and $P_0\approx0.6 \times10^{-12}$\,C/m for TMD bilayers \cite{Weston2022}). The in-plane relaxation displacement fields were calculated  as explained in Ref. \cite{Enaldiev_PRL} in terms of Fourier series $\bm{u}_{t/b}(\bm{r})=\sum_{n}\widetilde{u}_{\bm{g}_{n}}e^{i\bm{g}_n\bm{r}}.$ 

\begin{figure*}
	\centering
	\includegraphics[width=0.3\columnwidth]{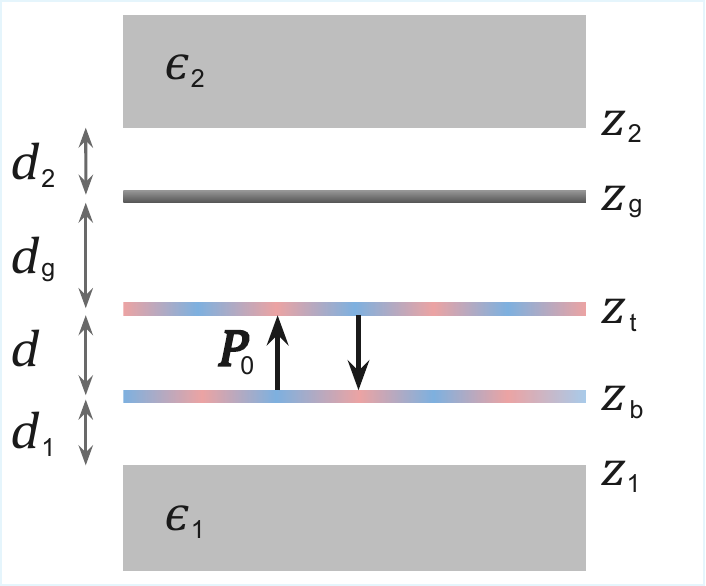}
	\caption{(a)  \label{fig:electrostatics} Sketch of the electrostatic model. In the limit $\epsilon_1\gg1$ the bottom medium simulates proximity gate electrode that doubles electrostatic moir\'e potential in graphene layer (at $z=z_g$).  }
\end{figure*}

In the electrostatic model we suppose that ferroelectric charges, producing the polarization in domains, sit in the middle plane of the layers such that the ferroelectric charge density is
\begin{align}
    n_{\rm ferro}(\bm{r},z)&= \frac{P(\bm{r})}{d}\left[\delta(z-z_t) -\delta(z-z_b)\right], 
\end{align}
where $d$ is interlayer distance in the bilayers. In addition to the ferroelectric charges we take into account the piezoelectric charges emerging due to  inhomogeneity of in-plane strain induced by the lattice relaxation  \cite{Enaldiev_PRL,Magorrian2021}. The piezoelectric charge density reads as follows  
\begin{equation}
    n_{\rm piezo}(\bm{r},z) = e_{11}\sum_{l=t,b}\left[2\partial_x u^{(l)}_{xy}+\partial_y(u^{(l)}_{xx}-u^{(l)}_{yy})\right]\delta(z-z_{l}),
\end{equation}
where $e_{11}$ is parameter of piezoelectric tensor \cite{Duerloo2012,Zhu2014}. Thus, electrostatic moir\'e potential results from solution of the Poisson equation:
\begin{gather}\label{Eq:Poisson}
\nabla\cdot\left[\epsilon(z)\nabla\varphi\left(\bm{r},z\right)\right]=-4\pi e\left[\left(n^{b}_{\rm fp}+n^b_{\rm ind}\right)\delta\left(z-z_{b}\right)+\left(n^{t}_{\rm fp}+n^t_{\rm ind}\right)\left(\bm{r}\right)\delta\left(z-z_{t}\right)+n^{gr}_{\rm ind}\left(\bm{r}\right)\delta\left(z-z_{g}\right)\right]. \\
\epsilon(z)=\left\{
\begin{array}{cc}
     \epsilon_2, &\quad z>z_2 \\
      1, &\quad z_1 < z < z_2 \\
      \epsilon_1, & z<z_1
      \end{array}
\right.
\end{gather}
Here, $\nabla=(\partial/\partial\bm{r},\partial/\partial z)$ and 2D densities are defined as follows:
\begin{gather}\label{Eq:ntb}
n^{t/b}_{\rm fp}=\pm\frac{P\left(\bm{r}\right)}{d}+e_{11}\left[2\partial_x u^{(t/b)}_{xy}+\partial_y(u^{(t/b)}_{xx}-u^{(t/b)}_{yy})\right],\\
n^{t/b}_{\rm ind} = -\alpha\left(\frac{\partial^2}{\partial x^2}+\frac{\partial^2}{\partial y^2}\right)\varphi\left(\bm{r},z_{t/b}\right) \label{Eq:nind} \\
n^{gr}_{\rm ind}\left(\bm{r}\right)=e\int d^3\bm{r'}\Pi\left(\bm{r}-\bm{r'},k_F\right)\varphi\left(\bm{r'},z_{gr}\right).\label{Eq:ngrind}
\end{gather}
Equations \eqref{Eq:nind} and \eqref{Eq:ngrind} describe induced polarization charges in t/b layer and graphene, respectively, which are determined by 2D polarizability, $\alpha=d(\epsilon_{||}-1)/4\pi$ (where $\epsilon_{||}$ is in-plane dielectric permittivity of bulk TMD/hBN crystal), and polarization operator of massless Dirac fermions in graphene, $\Pi(\bm{r}-\bm{r}'
,k_F)$ with the Fermi-wave number determined as $k_F=\sqrt{\pi n}$ ($n$ is carrier concentration) \cite{CastroNeto2009}. Looking for solution of Eq. \eqref{Eq:Poisson} in terms of a Fourier series $\varphi\left(\bm{r},z\right) = \sum_{\bm{g}_n} \left( \varphi_{n}(z) e^{i\bm{g}_n\bm{r}} + \text{c.c.} \right)$ with continuity boundary conditions for potential and corresponding break for out-of-plane component of electric field at each interface we obtain the following expression for the potential harmonic in graphene layer: 
\begin{equation}\label{Eq:phig}
	\varphi_n(z_g) = 
\frac{\frac{4\pi}{g_n}\left\{n^t_{n}\left[\Theta_n(d+d_1,\epsilon_1)+4\pi\alpha g_n\sh(g_nd)\Theta_n(d_1,\epsilon_1)\right]+n^b_{n}\Theta_n(d_1,\epsilon_1)\right\}\Theta_n(d_2,\epsilon_2)}
{\left|
            \begin{matrix}
		\Phi_n(d_2,\epsilon_2)+\frac{4\pi e^2}{g_n}\Pi(g_n,k_F)\Theta_n(d_2,\epsilon_2)	&  \Xi_n		\\
				-\Theta_n(d_2,\epsilon_2)	& \Lambda_n
			\end{matrix}
            \right|}
\end{equation} 
where $n^{t/b}_n$ is corresponding Fourier harmonic of the 2D charge density in t/b layer \eqref{Eq:ntb} and we introduced the following functions:
\begin{gather*}
	\Theta_n(d,\epsilon) = \ch\left(g_nd\right)+\epsilon\, \sh\left(g_nd\right),\nonumber \\
	\Phi_n(d,\epsilon) = \epsilon\, \ch\left(g_nd\right)+ \sh\left(g_nd\right), \nonumber \\
    \Xi_n\!=\!\Phi_n(d\!+\!d_1\!+\!d_g,\epsilon_1)\!+\!4\pi \alpha g_n\ch\left(g_n(d+d_g)\right)\Theta_n(d_1,\epsilon_1)\! +\! 4\pi\alpha g_n\ch\left(g_nd_g\right)\left[\Theta_n(d+d_1,\epsilon_1)\!+\!4\pi\alpha g_n\sh(g_nd)\Theta_n(d_1,\epsilon_1)\right],\\
    \Lambda_n\!=\!\Theta_n(d\!+\!d_1\!+\!d_g,\epsilon_1)\!+\!4\pi\alpha g_n\sh\left(g_n(d+d_g)\right)\Theta_n(d_1,\epsilon_1)\! +\! 4\pi\alpha g_n\sh\left(g_nd_g\right)\left[\Theta_n(d+d_1,\epsilon_1)\!+\!4\pi\alpha g_n\sh(g_nd)\Theta_n(d_1,\epsilon_1)\right].
\end{gather*}
In Table~\ref{tab:potentials} we gathered values of the potential harmonics values for graphene/twisted hBN bilayer heterostructure with $\theta=0.18^{\circ}$ and different electron concentrations.

\begin{table}[ht]
\centering
\caption{Non-zero harmonics of electrostatic moir\'e potential in graphene layer of graphene/twisted hBN heterostructure studied in the main text $V_{n}^{(a)}=-|e|\varphi_n(z_g)$ (in $\text{meV}$) at three values of electron concentration (in cm$^{-2}$). The value were calculated at the following parameters $a=2.5~{\rm \AA}$, $d = 3.35~{\rm \AA};\,d_1 = 3.35~{\rm \AA};\,d_2 = 3.34~{\rm \AA};\,d_g = 3.325~{\rm \AA}$, $\epsilon_1=\infty$, $\epsilon_2=3.5$ and $\alpha\approx0.93\,{\rm \AA}$. Due to $120^{\circ}$-rotational symmetry of potential, harmonics $\varphi_{n}$ for reciprocal vectors with the same magnitude (i.e. $\bm{g}^{(n)}_{2,3}=\hat{R}_z(\pm\frac{2\pi}{3})\bm{g}_{1}^{(n)}$; $\hat{R}_z(\pm\frac{2\pi}{3})$ is $\pm120^{\circ}$-rotation matrix) are equal to that of $\bm{g}^{(n)}_{1}$. Here $\bm{g}^{(1)}_1=(4\pi\theta/a\sqrt{3})(1/2,\sqrt{3}/2)$. }
\label{tab:potentials}
\begin{tabular}{cccc}
\hline
$g_n$ & $10^{10} ~\text{cm}^{-2}$ & $10^{11} ~\text{cm}^{-2}$ & $10^{12} ~\text{cm}^{-2}$ \\
\hline
$\bm{g}^{(1)}_1$ & 42.875 & 31.946 & 16.533 \\
$\bm{g}_1^{(3)}=2\bm{g}^{(1)}_1$ & 13.59 & 11.998 & 6.589 \\
$\bm{g}_1^{(6)}=3\bm{g}^{(1)}_1$ & 5.598 & 5.219 & 3.202 \\
$\bm{g}^{(10)}_{1}=4\bm{g}^{(1)}_1$ & 2.49 & 2.356 & 1.564 \\
$\bm{g}^{(15)}_{1}=5\bm{g}^{(1)}_1$ & 1.11 & 1.059 & 0.756 \\
$\bm{g}^{(21)}_{1}=6\bm{g}^{(1)}_1$ & 0.459 & 0.441 & 0.35 \\
$\bm{g}^{(29)}_{1}=7\bm{g}^{(1)}_1$ & 0.167 & 0.161 & 0.139 \\
$\bm{g}^{(40)}_{1}=8\bm{g}^{(1)}_1$ & 0.042 & 0.041 & 0.037 \\
$\bm{g}^{(50)}_{1}=9\bm{g}^{(1)}_1$ & 0.004 & 0.003 & 0.002 \\
\hline
\end{tabular}
\end{table}

\newpage
\section{Effective Hamiltonian for primary massless Dirac fermions}

To obtain Hamiltonian $H_{\bm{\gamma}}$ (Eq. (3) in the main manuscript) we, first, project out Hamiltonian $\hat{H}_0=v\bm{\sigma}\cdot\hat{\bm{p}}+V(\bm{r})$ on a set of plane waves corresponding to the  lowest Fourier harmonics ($\bm{g}_{1,2,3}\equiv\bm{g}_{1,2,3}^{(1)}$) of the moir\'e potential: $\psi\approx e^{i\bm{k}\bm{r}}\left(\widetilde{\psi}_0+\sum_{l=1,2,3}\left[\widetilde{\psi}_{\bm{g}_l}e^{i\bm{g}_l\bm{r}}+\widetilde{\psi}_{-\bm{g}_l}e^{-i\bm{g}_l\bm{r}}\right]\right)$, leading to 
\begin{multline}\label{Eq:Hgamma1}
      \begin{pmatrix}
  \hbar v\bm{\sigma}\cdot\bm{k}&  -\frac{V_1^{(a)}}{2i}\sigma_0 & \frac{V_1^{(a)}}{2i}\sigma_0 & -\frac{V_1^{(a)}}{2i}\sigma_0&  \frac{V_1^{(a)}}{2i}\sigma_0 & -\frac{V_1^{(a)}}{2i}\sigma_0 & \frac{V_1^{(a)}}{2i}\sigma_0\\
 \frac{V_1^{(a)}}{2i}\sigma_0  & \hbar v\bm{\sigma}\cdot(\bm{k}+\bm{g}_1) & 0 & 0 & 0&0 & 0  \\
 -\frac{V_1^{(a)}}{2i}\sigma_0 &0&\hbar v\bm{\sigma}\cdot(\bm{k}-\bm{g}_1)& 0& 0&0& 0\\
  \frac{V_1^{(a)}}{2i}\sigma_0 &0&0&\hbar v\bm{\sigma}\cdot(\bm{k}+\bm{g}_2) & 0 & 0 &0\\
 -\frac{V_1^{(a)}}{2i}\sigma_0&0&0&0&\hbar v\bm{\sigma}\cdot(\bm{k}-\bm{g}_2)&0&0\\
  \frac{V_1^{(a)}}{2i}\sigma_0 & 0 &0& 0 & 0&\hbar v\bm{\sigma}\cdot(\bm{k}+\bm{g}_3)&0\\
  -\frac{V_1^{(a)}}{2i}\sigma_0 &0 & 0 & 0 & 0 & 0 &\hbar v\bm{\sigma}\cdot(\bm{k}-\bm{g}_3)
    \end{pmatrix}
    \begin{pmatrix}
        \widetilde{\psi}_0\\
        \widetilde{\psi}_{\bm{g}_1}\\
        \widetilde{\psi}_{-\bm{g}_1}\\
        \widetilde{\psi}_{\bm{g}_2}\\
        \widetilde{\psi}_{-\bm{g}_2}\\
        \widetilde{\psi}_{\bm{g}_3}\\
        \widetilde{\psi}_{-\bm{g}_3}\\
    \end{pmatrix}=\\
    =\varepsilon\begin{pmatrix}
        \widetilde{\psi}_0\\
        \widetilde{\psi}_{\bm{g}_1}\\
        \widetilde{\psi}_{-\bm{g}_1}\\
        \widetilde{\psi}_{\bm{g}_2}\\
        \widetilde{\psi}_{-\bm{g}_2}\\
        \widetilde{\psi}_{\bm{g}_3}\\
        \widetilde{\psi}_{-\bm{g}_3}\\
    \end{pmatrix}.
\end{multline}
where we take into account only matrix elements between $\widetilde{\psi}_{0}$ and  $\widetilde{\psi}_{\bm{g}_l\neq 0}$, whereas those between $\widetilde{\psi}_{\bm{g}_l\neq 0}$ were neglected. For $\bm{k}=0$ two eigen states of matrix \eqref{Eq:Hgamma1}
possesses zero energy:
\begin{align}
    \widetilde{\psi}_{+1}&=\frac{1}{\sqrt{6+\left(\frac{2\hbar v |\bm{g}_1|}{V_1^{(a)}}\right)^2}}\left(\frac{2\hbar v |\bm{g}_1|}{V_1^{(a)}},0,0,e^{i\frac{2\pi}{3}},0,e^{i\frac{2\pi}{3}},0,-e^{i\frac{\pi}{3}},0,-e^{i\frac{\pi}{3}},0,1,0,1\right)^{\rm T},\\
    \widetilde{\psi}_{-1}&=\frac{1}{\sqrt{6+\left(\frac{2\hbar v |\bm{g}_1|}{V_1^{(a)}}\right)^2}}\left(0,\frac{2\hbar v |\bm{g}_1|}{V_1^{(a)}},e^{-i\frac{2\pi}{3}},0,-e^{i\frac{\pi}{3}},0,e^{i\frac{2\pi}{3}},0,e^{i\frac{2\pi}{3}},0,1,0,1,0\right)^{\rm T}.
\end{align}

Projecting the Hamiltonian with $\bm{k}\neq0$ onto subspace of the zero energy states: $\psi=C_1(\bm{k})\widetilde{\psi}_{+1}+C_2(\bm{k})\widetilde{\psi}_{-1}$, we arrive to the following effective model:
\begin{equation}\label{Eq:Hgamma2}
    \begin{pmatrix}
    0& \frac{\hbar v}{1+\frac{3}{2}\left(\frac{V_1^{(a)}}{\hbar v|\bm{g}_1|}\right)^2} (k_x-ik_y)\\
    \frac{\hbar v}{1+\frac{3}{2}\left(\frac{V_1^{(a)}}{\hbar v|\bm{g}_1|}\right)^2}(k_x+ik_y) & 0
    \end{pmatrix}
    \begin{pmatrix}
        C_1\\
        C_2
    \end{pmatrix}=\varepsilon\begin{pmatrix}
        C_1\\
        C_2
    \end{pmatrix}.
\end{equation}
Substituting $|\bm{g}_1|=4\pi\theta/a\sqrt{3}$ in the left-hand side of Eq. \eqref{Eq:Hgamma2} we obtain $\hat{H}_{\bm{\gamma}}$ from the main text. 

\section{Effective Hamiltonian for secondary massless Dirac fermions }

Secondary Dirac fermions emerge in vicinity of $\bm{k}=\bm{g}_{1,2,3}/2=\bm{\mu}_{1,2,3}$. Due to $120^{\circ}$-rotational symmetry it is enough to derive effective Hamiltonian for $\bm{\mu}_3=(g_1/2,0)$, that can be obtained leaving only the first and last lines in the left-hand side of Eq. \eqref{Eq:Hgamma1} (i.e. approximating $\psi\approx e^{i\bm{k}\bm{r}}\left(\widetilde{\psi}_0+\widetilde{\psi}_{\bm{g}_3}e^{-i\bm{g}_3\bm{r}}\right)$) and putting $\bm{k}=\bm{q}+\bm{g}_3/2$. This leads to:
\begin{equation}
    \hat{H}=
    \begin{pmatrix}
        \frac{\hbar v g_1}{2}\sigma_x +\hbar v\bm{\sigma}\cdot\bm{q} & \frac{V_1^{(a)}}{2i}\sigma_0 \\
        \frac{V_1^{(a)}}{2i}\sigma_0 &   -\frac{\hbar v g_1}{2}\sigma_x+\hbar v\bm{\sigma}\cdot\bm{q}
    \end{pmatrix}.
\end{equation}
Applying unitary transformation $\hat{U}$ to $\hat{H}$ we obtain 
\begin{equation}\label{Eq:Ham_mu2}
    \hat{\widetilde{H}}=\hat{U}\hat{H}\hat{U}^{+}=
\begin{pmatrix}
\varepsilon_{\bm{\mu}}+\hbar v\sigma_zq_y \frac{V_1^{(a)}}{2\varepsilon_{\bm{\mu}}}-\hbar v \sigma_xq_x &  i\frac{\hbar v g_1}{2\varepsilon_{\bm{\mu}}}\hbar v\sigma_zq_y \\
-i\frac{g_1}{2\varepsilon_{\bm{\mu}}}\sigma_zq_y &  -\varepsilon_{\bm{\mu}}-\hbar v\sigma_zq_y \frac{V_1^{(a)}}{2\varepsilon_{\bm{\mu}}}-\hbar v \sigma_xq_x 
\end{pmatrix}
\end{equation}
where
\begin{equation}
            \hat{U}=
           \frac{1}{2} \begin{pmatrix}
                1 & \frac{1}{2}\frac{\hbar vg_1-iV_1^{(a)}}{\varepsilon_{\bm{\mu}}} & \frac{1}{2}\frac{\hbar vg_1-iV_1^{(a)}}{\varepsilon_{\bm{\mu}}} & -1\\
               -\frac{1}{2}\frac{\hbar vg_1-iV_1^{(a)}}{\varepsilon_{\bm{\mu}}}  & -1 & 1 & -\frac{1}{2}\frac{\hbar vg_1-iV_1^{(a)}}{\varepsilon_{\bm{\mu}}}\\
                 1 &-\frac{1}{2}\frac{\hbar vg_1-iV_1^{(a)}}{\varepsilon_{\bm{\mu}}} & -\frac{1}{2}\frac{\hbar vg_1-iV_1^{(a)}}{\varepsilon_{\bm{\mu}}}&-1 \\
        \frac{1}{2}\frac{\hbar vg_1-iV_1^{(a)}}{\varepsilon_{\bm{\mu}}}         & -1& 1 & \frac{1}{2}\frac{\hbar vg_1-iV_1^{(a)}}{\varepsilon_{\bm{\mu}}}
            \end{pmatrix}
\end{equation}
and $\varepsilon_{\bm{\mu}}=(1/2)\sqrt{(\hbar v g_1)^2+\left(V_1^{(a)}\right)^2}$.
Next, we consider eigenvalue problem for Hamiltonian \eqref{Eq:Ham_mu2}:
\begin{equation}
    \begin{pmatrix}
\varepsilon_{\bm{\mu}}+\hbar v\sigma_zq_y \frac{V_1^{(a)}}{2\varepsilon_{\bm{\mu}}}-\hbar v \sigma_xq_x &  i\frac{\hbar v g}{2\varepsilon_{\bm{\mu}}}\hbar v\sigma_zq_y \\
-i\frac{g}{2\varepsilon_{\bm{\mu}}}\sigma_zq_y &  -\varepsilon_{\bm{\mu}}-\hbar v\sigma_zq_y \frac{V_1^{(a)}}{2\varepsilon_{\bm{\mu}}}-\hbar v \sigma_xq_x 
\end{pmatrix}\begin{pmatrix}
    \psi_{+}\\
    \psi_-
\end{pmatrix} = \varepsilon \begin{pmatrix}
    \psi_{+}\\
    \psi_-
\end{pmatrix}
\end{equation}
and express $\psi_-$ via $\psi_+$ ($\psi_+$ via $\psi_-$) for states characterized by energies in vicinity of secondary Dirac point $\varepsilon\approx\varepsilon_{\bm{\mu}}$ ($\varepsilon\approx-\varepsilon_{\bm{\mu}}$). This leads us to the following equation:
\begin{equation}\label{Eq:Hmu3}
\left[\pm\varepsilon_{\bm{\mu}}\pm\hbar v\sigma_zq_y \frac{V_1^{(a)}}{2\varepsilon_{\bm{\mu}}}\pm \frac{(\hbar v g)^2}{8\varepsilon^3_{\bm{\mu}}}\left(\hbar v q_y\right)^2-\hbar v \sigma_xq_x\right]\psi_{\pm}=\varepsilon\psi_{\pm}.
\end{equation}
After exchange $v\to v_*$ in the left-hand side of Eq. \eqref{Eq:Hmu3} (which, as we checked, gives better approximation to exact dispersion) we arrive to the Hamiltonian for secondary Dirac fermions introduced in the main text.  

\newpage
\section{Miniband structure in graphene/twisted ferroelectric bilayer at finite out-of-plane electric field across polar domains}
\begin{figure*}[h!]
	\includegraphics[width=0.6\columnwidth]{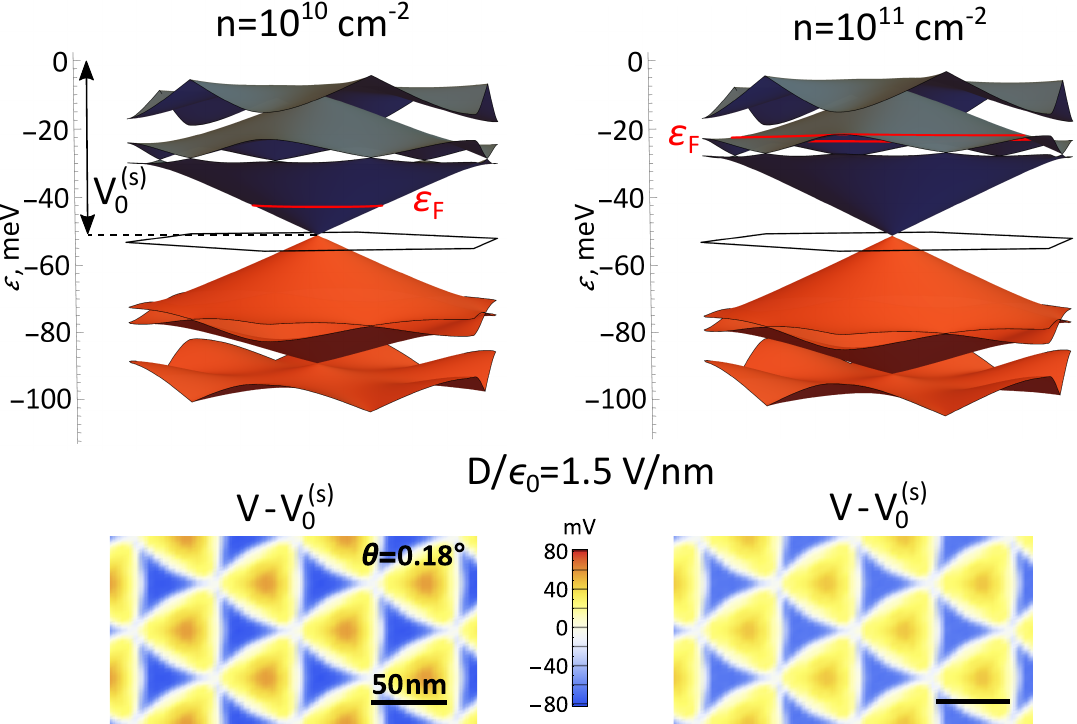}
	\caption{(a)  \label{fig:minibandsSM} Mini-band structures in graphene/twisted hBN bilayer at out-of-plane electric field $D/\epsilon_0=1.5$\, V/nm across the twisted interface with $\theta=0.18^{\circ}$. Bottom inset show polar domain structure for the electric field.}
\end{figure*}

\section{Resonant absorption and shift photocurrent at finite out-of-plane electric field across twisted interface}

\begin{figure*}[h!]
	\includegraphics[width=\columnwidth]{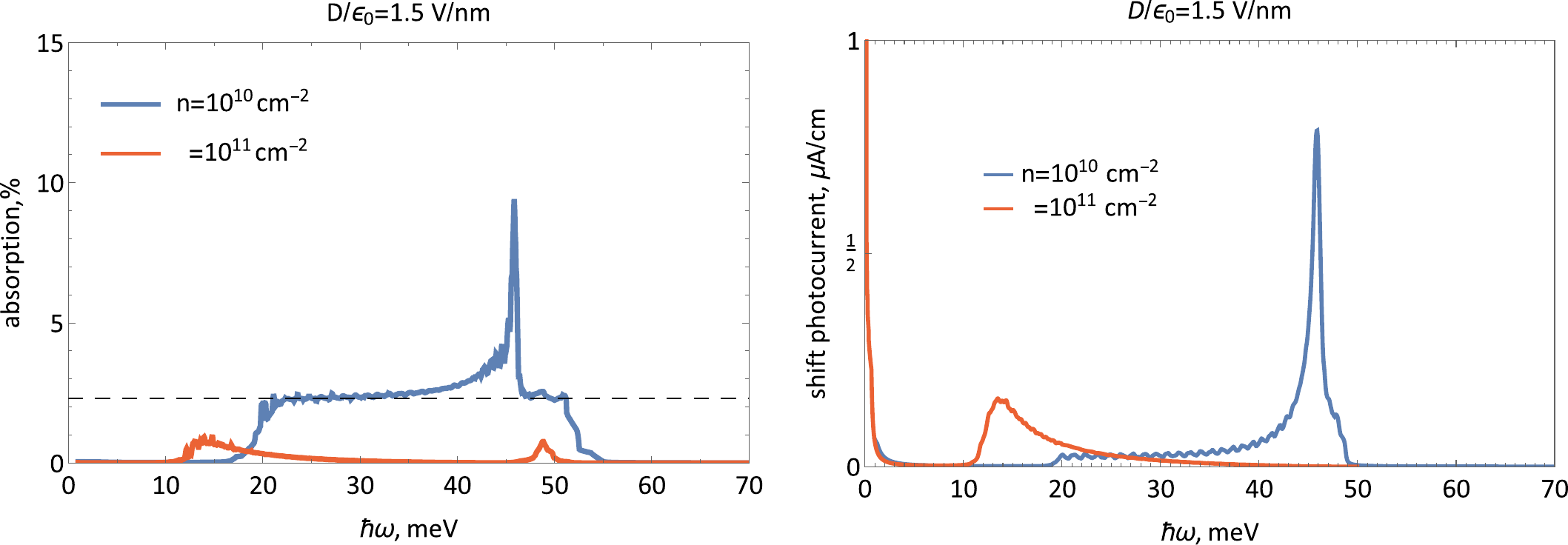}
	\caption{(a)  \label{fig:abs&jSM} Frequency dependences of absorption (left) and shift photocurrent (right) for graphene/twisted hBN bilayer heterostructure at out-of-plane electric field $D/\epsilon_0=1.5$\, V/nm across the twisted interface and $\theta=0.18^{\circ}$. The shift photocurrent was calculated at the intensity of incident radiation, $I=0.1$\,W/cm$^{2}$. Dashed line on the left panel shows absorption of isolated graphene $\approx 2.3\%$.}
\end{figure*}

\newpage
\section{Derivation of the injection and shift photocurrents}

In this section we derive equations (8)-(10) in the main text. For that we consider the following Hamiltonian:
\begin{align}
    \hat{H}&=\hat{H}_0 + \hat{H}_1(t),\\
    \hat{H}_0 &=v\bm{\sigma}\cdot\bm{\hat{p}} + V(\bm{r}),\\
    \hat{H}_1(t) &=v\bm{\sigma}\cdot\bm{A}(t)=\frac{A_0}{2}v\bm{\sigma}\cdot\bm{\eta}\left(e^{i\omega t+\gamma t}+e^{-i\omega t+\gamma t}\right),
\end{align}
where we keep designations introduced in the main text and add $\gamma>0$ ($\gamma\ll\omega$) assuming vanishing of the wave at $t\to-\infty$ . Then, we look for density matrix as a series  $\hat{\rho}=\hat{\rho}_0 + \hat{\rho}_1 + \hat{\rho}_2$, where $\hat{\rho}_0=\sum_{n,\bm{k}}f(\varepsilon_{n}(\bm{k}))|n\bm{k}\rangle\langle n\bm{k}|$ is equilibrium density matrix ($\langle\bm{r}|n\bm{k}\rangle\equiv\psi_{n\bm{k}}(\bm{r})$ is Bloch state), while $\hat{\rho}_1\propto A_0$ and $\hat{\rho}_2\propto A_0^2$. To determine the first and second amendments to equilibrium density matrix we perturbatively solve Liouville equation:  
\begin{equation}\label{Eq:LE}
    \frac{\partial(\hat{\rho}_1 + \hat{\rho}_2)}{\partial t} = \frac{i}{\hbar} \big[\hat{H}_0 + \hat{H}_1,\, \hat{\rho}_0 + \hat{\rho}_1 + \hat{\rho}_2 \big].
\end{equation}

From Eq. \eqref{Eq:LE} we find
\begin{equation}\label{Eq:rho2}
    \hat{\rho}_2 =   \frac{1}{\left(i\hbar\right)^2}  \int_{-\infty}^{t} dt' \int_{-\infty}^{t'} dt'' e^{\frac{-i }{\hbar}\hat{H}_0 (t-t')} \left[ \hat{H}_1(t'), e^{\frac{-i }{\hbar}\hat{H}_0 (t'-t'')} [\hat{H}_1(t''), \hat{\rho}_0] e^{\frac{i }{\hbar}\hat{H}_0 (t'-t'')} \right] e^{\frac{i}{\hbar}\hat{H}_0 (t-t')}.
\end{equation}
Matrix element on the Bloch states for zero-frequency part of density operator  \eqref{Eq:rho2} reads as follows:
\begin{multline}\label{Eq:rho2_me}
    \langle n'\bm{k}|\hat{\rho}_2|n\bm{k}\rangle =\frac{e^2A_0^2}{4}\sum_{\substack{\Omega=\pm \hbar\omega\\n''}}\frac{(\bm{v}_{n'n''}(
    \bm{k})\cdot\bm{\eta})(\bm{v}_{n''n}(
    \bm{k})\cdot\bm{\eta})}{\varepsilon_{n'}(\bm{k})-\varepsilon_n(\bm{k})-i\gamma}\left[\frac{f(\varepsilon_n(\bm{k}))-f(\varepsilon_{n''}(\bm{k}))}{\varepsilon_{n''}(\bm{k})-\varepsilon_n(\bm{k})-\Omega-i\gamma}+\frac{f(\varepsilon_{n'}(\bm{k}))-f(\varepsilon_{n''}(\bm{k}))}{\varepsilon_{n'}(\bm{k})-\varepsilon_{n''}(\bm{k})-\Omega-i\gamma}\right]    
\end{multline}
where  $\bm{v}_{nn'}(\bm{k})=\langle n\bm{k}|v\bm{\sigma}|n'\bm{k}\rangle$ is velocity operator matrix element.

Calculating current with the help of Eq. \eqref{Eq:rho2_me} we obtain:
\begin{align}\label{Eq:gen_j}
    \bm{j}&=\frac{e}{S}\sum_{n,\bm{k}}\langle n\bm{k}| v\bm{\sigma}\hat{\rho}_2|n\bm{k}\rangle = \frac{e}{S}\sum_{n,n',\bm{k}}\langle n\bm{k}| v\bm{\sigma}|n'\bm{k}\rangle\langle n'\bm{k}|\hat{\rho}_2|n\bm{k}\rangle = \nonumber\\
    &=\frac{e^3A_0^2}{4S}\sum_{\substack{\Omega=\pm \hbar\omega\\n'',n',n,\bm{k}}}\frac{\bm{v}_{nn'}(
    \bm{k})(\bm{v}_{n'n''}(
    \bm{k})\cdot\bm{\eta})(\bm{v}_{n''n}(
    \bm{k})\cdot\bm{\eta})}{\varepsilon_{n'}(\bm{k})-\varepsilon_n(\bm{k})-i\gamma}\left[\frac{f(\varepsilon_n(\bm{k}))-f(\varepsilon_{n''}(\bm{k}))}{\varepsilon_{n''}(\bm{k})-\varepsilon_n(\bm{k})-\Omega-i\gamma}+\frac{f(\varepsilon_{n'}(\bm{k}))-f(\varepsilon_{n''}(\bm{k}))}{\varepsilon_{n'}(\bm{k})-\varepsilon_{n''}(\bm{k})-\Omega-i\gamma}\right] \nonumber\\
    &=\frac{e^3A_0^2}{4S}\sum_{\substack{\Omega=\pm\hbar\omega\\n'',n'=n,\bm{k}}}\frac{\bm{v}_{nn}(
    \bm{k})(\bm{v}_{nn''}(
    \bm{k})\cdot\bm{\eta})(\bm{v}_{n''n}(
    \bm{k})\cdot\bm{\eta})}{-i\gamma}\left(\frac{f(\varepsilon_n(\bm{k}))-f(\varepsilon_{n''}(\bm{k}))}{\varepsilon_{n''}(\bm{k})-\varepsilon_n(\bm{k})-\Omega-i\gamma}-\frac{f(\varepsilon_n(\bm{k}))-f(\varepsilon_{n''}(\bm{k}))}{\varepsilon_{n''}(\bm{k})-\varepsilon_{n}(\bm{k})-\Omega+i\gamma}\right)+\nonumber \\
    &+\frac{e^3A_0^2}{4S}\sum_{\substack{\Omega=\pm \hbar\omega\\n'',n'\neq n,\bm{k}}}\frac{f(\varepsilon_n(\bm{k}))-f(\varepsilon_{n''}(\bm{k}))}{\varepsilon_{n'}(\bm{k})-\varepsilon_n(\bm{k})}\left[\frac{\bm{v}_{nn'}(
    \bm{k})(\bm{v}_{n'n''}(
    \bm{k})\cdot\bm{\eta})(\bm{v}_{n''n}(
    \bm{k})\cdot\bm{\eta})}{\varepsilon_{n''}(\bm{k})-\varepsilon_n(\bm{k})-\Omega-i\gamma}+\frac{\bm{v}_{nn'}^*(
    \bm{k})(\bm{v}_{n'n''}^*(
    \bm{k})\cdot\bm{\eta})(\bm{v}_{n''n}^*(
    \bm{k})\cdot\bm{\eta})}{\varepsilon_{n''}(\bm{k})-\varepsilon_{n}(\bm{k})-\Omega+i\gamma}\right]
\end{align}
Applying Sokhotski-Plemelj formula and replacing $\gamma\to\hbar/\tau$ in the first term ($n'=n$) after the last equality in Eq. \eqref{Eq:gen_j} we obtain the expression for the injection photocurrent in the main manuscript:
\begin{multline}
    \bm{j}_{\rm inj}=\frac{e^3A_0^2\pi\tau}{2\hbar S}\sum_{\substack{\Omega=\pm \hbar\omega\\n'',n,\bm{k}}}\bm{v}_{nn}(
    \bm{k})(\bm{v}_{nn''}(
    \bm{k})\cdot\bm{\eta})(\bm{v}_{n''n}(
    \bm{k})\cdot\bm{\eta})\left[f(\varepsilon_{n''}(\bm{k}))-f(\varepsilon_{n}(\bm{k}))\right]\delta\left(\varepsilon_{n''}(\bm{k})-\varepsilon_n(\bm{k})-\Omega\right),
\end{multline}
whereas the second term transforms into the expression for the shift photocurrent by virtue of $\bm{v}_{n'n}(\bm{k})=-e^{i\alpha_n(\bm{k})-i\alpha_{n'}(\bm{k})}\bm{v}_{n'n}^*(-\bm{k})$ and $\varepsilon_{n}(-\bm{k})=\varepsilon_n(\bm{k})$ imposed by the hidden symmetry $\hat{T}=i\sigma_y\hat{K}$ and taking the limit $\gamma\to+0$. This leads to 
\begin{multline}
  \bm{j}_{\rm shift} =  \frac{e^3A_0^2\pi}{2S}\sum_{\substack{\Omega=\pm \hbar\omega\\n'',n'\neq n,\bm{k}}}\frac{f(\varepsilon_n(\bm{k}))-f(\varepsilon_{n''}(\bm{k}))}{\varepsilon_{n'}(\bm{k})-\varepsilon_n(\bm{k})}\delta\left(\varepsilon_{n''}(\bm{k})-\varepsilon_n(\bm{k})-\Omega\right){\rm Im}\left[\bm{v}_{nn'}(
    \bm{k})(\bm{v}_{n'n''}(
    \bm{k})\cdot\bm{\eta})(\bm{v}_{n''n}(
    \bm{k})\cdot\bm{\eta})\right].
\end{multline}

\end{widetext}
\end{document}